\documentclass[unnumsec,webpdf,modern,large,namedate]{oup-authoring-template}

\usepackage{setspace}

\graphicspath{{Fig/}}

\theoremstyle{thmstyleone}%
\theoremstyle{thmstyletwo}%
\theoremstyle{thmstylethree}%

\usepackage{color,soul}
\usepackage{upgreek}

\newcommand{\sm}{$\mathcal{S}$} 
\newcommand{\am}{$\mathcal{A}$} 

\begin{document}

\journaltitle{Microscopy and Microanalysis}
\DOI{DOI HERE}
\copyrightyear{2025}
\pubyear{2025}
\access{Advance Access Publication Date: Day Month 2025}
\appnotes{Paper}

\firstpage{1}


\title[Scattering matrix recon.]{Quantitative structure determination from experimental four-dimensional scanning transmission electron microscopy via the scattering matrix}

\author[1]{E.W.C. Terzoudis-Lumsden}
\author[1]{A. Sadri}
\author[2,3]{M. Weyland}
\author[2,3]{L. Bourgeois}
\author[4]{S.M. Ribet}
\author[4,5]{G. Varnavides}
\author[6]{C. Ophus}
\author[2]{T.C. Petersen}
\author[1,$\ast$]{S.D. Findlay\ORCID{0000-0003-4862-4827}}

\authormark{Terzoudis-Lumsden et al.}

\address[1]{\orgdiv{School of Physics and Astronomy}, \orgname{Monash University}, \orgaddress{\street{Melbourne}, \postcode{3800}, \state{VIC}, \country{Australia}}}
\address[2]{\orgdiv{Monash Centre for Electron Microscopy}, \orgname{Monash University}, \orgaddress{\street{Melbourne}, \postcode{3800}, \state{VIC}, \country{Australia}}}
\address[3]{\orgdiv{Department of Materials Science and Engineering}, \orgname{Monash University}, \orgaddress{\street{Melbourne}, \postcode{3800}, \state{VIC}, \country{Australia}}}
\address[4]{\orgdiv{National Center for Electron Microscopy}, \orgname{Lawrence Berkeley National Laboratory}, \orgaddress{\street{Berkeley}, \postcode{94720}, \state{CA}, \country{USA}}}
\address[5]{\orgdiv{Miller Institute for Basic Research in Science}, \orgname{University of California}, \orgaddress{\street{Berkeley}, \postcode{94720}, \state{CA}, \country{USA}}}
\address[6]{\orgdiv{Department of Materials Science and Engineering}, \orgname{Stanford University}, \orgaddress{\street{Stanford}, \postcode{94305}, \state{CA}, \country{USA}}}

\corresp[$\ast$]{Corresponding author. \href{email:scott.findlay@monash.edu}{scott.findlay@monash.edu}}

\received{Date}{0}{Year}
\revised{Date}{0}{Year}
\accepted{Date}{0}{Year}



\abstract{
Considerable inroads have recently been made on algorithms to determine the sample potential from four-dimensional scanning transmission electron microscopy data from thick samples where multiple scattering cannot be neglected. This paper further develops the scattering matrix approach to such structure determination. Through simulation, we demonstrate how this approach can be modified to better handle partial spatial coherence, unknown probe defocus, and information from the dark field region. By combining these developments we reconstruct the electrostatic potential of a monolithic SrTiO$_3$ crystal showing good quantitative agreement with the expected structure.
}


\maketitle

\singlespacing
\begin{large}
\section{\LARGE Introduction}
Over the past two decades, scanning transmission electron microscopy (STEM) has benefited greatly from advances in probe corrector and imaging detector technologies. We can now routinely generate sub-{\AA}ngstr\"{o}m probes that are largely aberration-free. Coupled with fast readout pixel detectors, this enables recording the full diffraction pattern at each scan position of an atomically-fine probe at raster speeds sufficient to largely overcome sample drift and beam damage. The resultant 2D array of 2D diffraction patterns is widely called 4D STEM \citep{ophus2019four}. These advances in hardware have led to corresponding advances in ways to process and analyse this new data. Examples include strain and crystal orientation mapping at the nanoscale becoming more routine \citep{ophus2019four,mukherjee2020lattice,ophus2022automated}, and the application of machine learning strategies to this wealth of scattering data \citep{li2019manifold,kalinin2022machine,munshi2022disentangling,roccapriore2022automated,kalinin2023machine,gleason2024random}.

There has also been much innovation in, and application of, phase contrast and structure determination methods such as differential phase contrast and ptychography (see \citet{varnavides2023iterative} and references therein). Initially, such techniques largely relied on the phase object approximation, limiting quantitative determination of the specimen potential to thin specimens (though qualitatively interpretable reconstructions may be possible in thicker samples, especially with correction strategies such as that of \citet{gao2022overcoming}). More recently, however, approaches have been developed capable of reconstructing the electrostatic potential from thicker samples, where the effects of multiple electron scattering can no longer be neglected. These include multislice-based methods  \citep{maiden2012ptychographic,van2012method,van2013general,ren2020multiple,chen2021electron,sha2022deep,diederichs2024exact,ribet2024uncovering,schloz2024improved} and scattering-matrix-based methods \citep{pennington2014third,wang2016inversion,brown2018,donatelli2020inversion,brown2020three,pelz2021phase,zeltmann2023uncovering}. In this paper, we further develop the scattering matrix approach.

Determining the scattering matrix from experimental measurements offers some interesting possibilities. For crystalline samples, \citet{findlay2021scattering} predicted that scattering matrix determination together with its symmetry properties could be used to estimate sample thickness without knowing the structure. Being the quantum mechanical operator describing wavefield scattering through a sample, an experimentally-determined scattering matrix would also enable post-synthesis of imaging geometries and modes distinct from the experiment performed, including phase-sensitive modes that provide information on the variation in structure along the beam direction and for which there is no direct experimental analogue \citep{ophus2019advanced,brown2020three,terzoudis2023resolution}. Moreover, scattering matrix determination offers a route to reconstructing the projected electrostatic potential of a periodic sample \citep{spence1999,allen1999,allen2000,findlay2005}. This was achieved on 4D STEM data from multiple defocus values by \citet{brown2018}, who used the algorithmically-robust (since it involves solving only linear equations) inversion method of \citet{allen1999,allen2000} to reconstruct the potential. Because requiring multiple defocus values complicates the experiment, \citet{findlay2021scattering} subsequently proposed an approach using data from only a single defocus. \citet{sadri2023determining} showed that reconstruction artefacts in the earlier inversion method, due to the probe-forming aperture constraining the extent of the scattering matrix that can be determined, can be overcome via an optimisation-based strategy.

These predictions of \citet{findlay2021scattering} and \citet{sadri2023determining} were, however, only shown through simulation. This paper provides the experimental validation. Achieving this required overcoming three obstacles associated with real experimental data. The first is the significant impact of partial spatial coherence. We better account for this effect using a modified phase retrieval algorithm following \citet{clark2011simultaneous}. The second is that the prediction of \citet{findlay2021scattering} for estimating thickness via scattering matrix symmetry neglects the impact of unknown lens aberrations. We show that when, as is often the case in experiment, the dominant unknown aberration is probe defocus then the symmetry constraint estimates the value of a linear combination of the unknown thickness and defocus --- not as good as an estimate for either individually, but a useful constraint nonetheless. The third is that \citet{sadri2023determining} only used 4D STEM data in the bright field region to reconstruct the sample potential. We show how to include dark-field information, which can improve reconstruction fidelity.

The paper is structured as follows. We first summarise our scattering-matrix-based approach to structure determination. We then present new developments within the reconstruction procedure --- handling partial spatial coherence, lens aberrations, and the dark field region --- and explore via simulation how this improves the reconstruction of the sample potential. Combining these advances, we demonstrate a successful reconstruction of the electrostatic potential of a monolithic SrTiO$_3$ sample from experimental 4D STEM data, and show it to be in close quantitative agreement with expectations.

\section{\LARGE Background and methods}\label{sec:background}

In this section we outline the underlying theory behind the forward problem for 4D STEM, as depicted in the top row of Fig. \ref{fig:intro_schematic}. We then outline our two-step approach to the inverse problem, first using measured 4D STEM intensities to reconstruct the scattering matrix following \citet{findlay2021scattering} and second using the reconstructed scattering matrix to solve for the structure following \citet{sadri2023determining}, as depicted in the bottom row of Fig. \ref{fig:intro_schematic}. We note a few minor modifications over those previous approaches. We also compare reconstructions on 4D STEM data simulated using two different models for thermal scattering.

\begin{figure*}[ht]
    \centering
    \includegraphics[width=\textwidth]{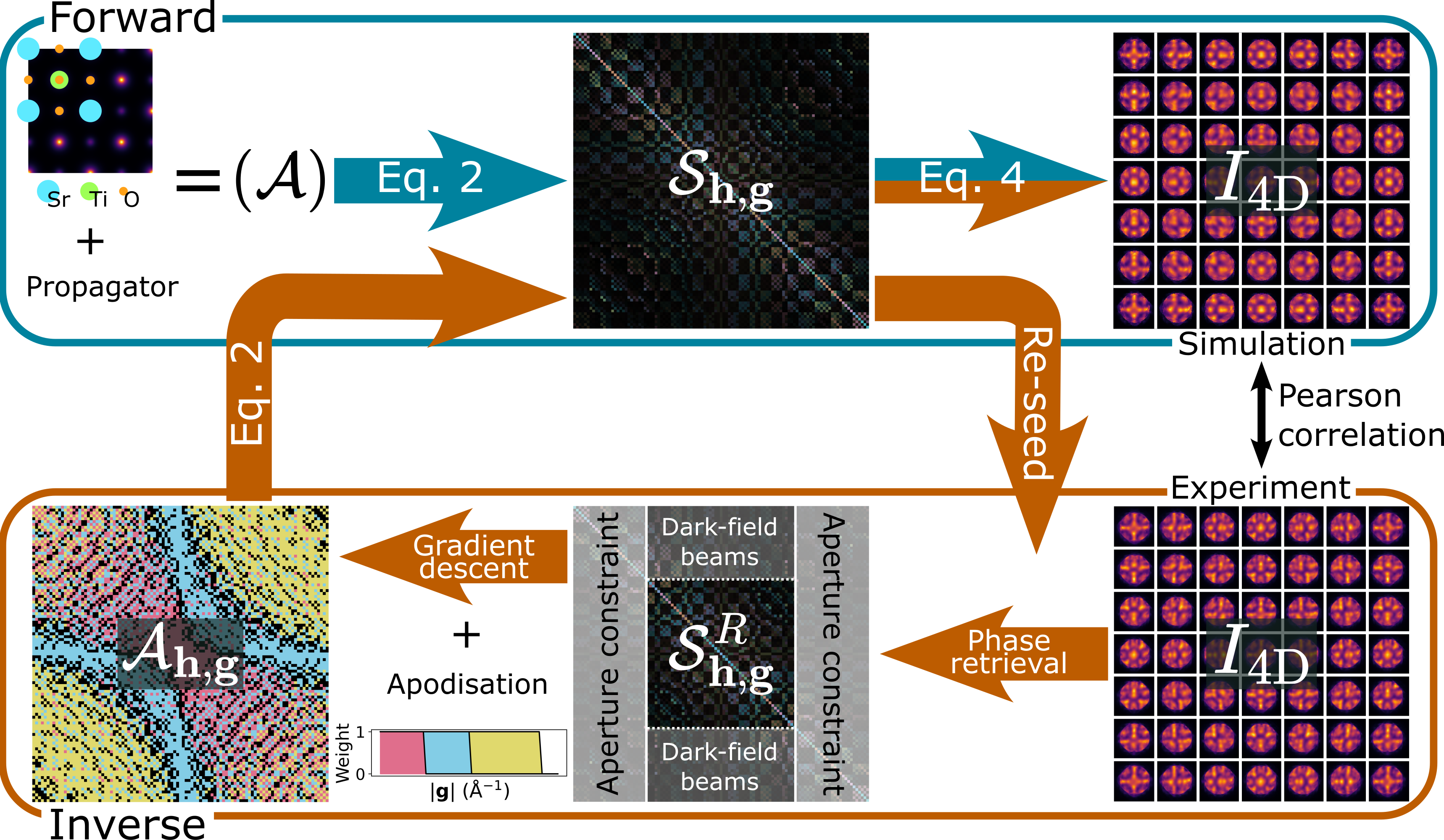}
    \caption{\emph{Top row}: Schematic outlining the forward problem. A known structure is encoded into the structure matrix {\am} from which the scattering matrix {\sm} and then the 4D STEM intensities $I_{\rm 4D}$ can be calculated. \emph{Bottom row}: Schematic outlining the inverse problem. From measured 4D STEM data $I_{\rm 4D}$, phase retrieval allows {\sm} to be partially reconstructed (some portions are inaccessible due to the probe-forming aperture). From the reconstructed {\sm}, gradient descent, with apodisation constraining the Fourier coefficients refined, solves for {\am} and thus the structure. Results can be improved by using the {\am} estimate to simulate a new {\sm} estimate to re-seed the phase retrieval step, and iterating over this process. Comparing simulated and measured 4D STEM intensities tracks progress and informs a stopping criterion.}
    \label{fig:intro_schematic}
\end{figure*}

\subsection{\Large The forward problem}
The Bloch wave formulation of fast electron scattering introduces the so-called structure matrix {\am}, comprising the electrostatic potential of the sample and geometric factors relating to the wavefield propagation \citep{humphreys1979scattering}:
\begin{equation}\label{eq:a_matrix}
    \mathcal{A}_{{\bf h},{\bf g}} = 
    \left \{ \begin{aligned}
    -({\bf k}_\perp+{\bf h})^2 &+ i U'_0, && \text{if}\ {\bf h} = {\bf g} \\
    U_{{\bf g}} &+ i U'_{{\bf g}}, && \text{otherwise}
  \end{aligned} \right. \;.
\end{equation}
Here, ${\bf k}_\perp = (k_x, k_y)$ denotes the transverse wavevector (i.e. the component of the mean wavevector in the convergent STEM probe perpendicular to the surface normal of the sample), the $U_{\bf g}$ denote the Fourier coefficients of the projected scattering potential,\footnote{More precisely, $U_{\bf g} = (2 m e / h^2) V_{\bf g}$, where the $V_{\bf g}$ are the Fourier coefficients of the electrostatic potential of the sample, $m$ is the relativistically corrected electron mass, $e$ is the elementary charge, and $h$ is Planck's constant.} and the $U'_{{\bf g}}$ denote the Fourier coefficients of the absorptive potential, here used to account for thermal scattering \citep{humphreys1979scattering,bird1990absorptive}.

The scattering of paraxial electron wavefields through a single crystal sample can be described via the so-called scattering matrix {\sm}, related to {\am} via
\begin{equation}\label{eq:a_to_s}
    \mathcal{S} = \exp{\left(i \frac{\pi t}{K} \mathcal{A}\right)} \;,
\end{equation}
where $t$ is the sample thickness and $K$ is the relativistically-corrected electron wavenumber. Specifically, the incident electron wavefunction, $\psi_{\rm in}$, and the exit surface wavefunction, $\psi_{\text{out}}$, expressed in a basis of plane waves (indexed by the transverse wavevectors ${\bf g}$ and ${\bf h}$ respectively), are related via
\begin{equation}\label{eq:s_reciprocal}
    \psi_{\text{out}}({\bf h}) = \sum_{{\bf g}} \mathcal{S}_{{\bf h}, {\bf g}} \psi_{\text{in}}({\bf g}) \;.
\end{equation} 

The incident electron wavefield in STEM may be written as $\psi_{\text{in}}({\bf g}) = T({\bf g}) e^{-2 \pi i {\bf g} \cdot {\bf R}}$, where the pre-specimen lens transfer function $T({\bf g}) = O({\bf g}) e^{-2 \pi i K \chi({\bf g})}$ (with aperture function\footnote{The aperture function is often denoted by $A$, but we have adopted a different notation to avoid confusion with the structure matrix {\am}.} $O({\bf g})$ and aberration function $\chi({\bf g})$), and where ${\bf R}$ denotes the STEM probe position on the sample. Using Eq. (\ref{eq:s_reciprocal}), the 4D STEM intensity is thus given by:
\begin{equation}\label{eq:4d_sm}
    I_{\rm 4D}({\bf h}, {\bf R}) = \left| \sum_{{\bf g}} \mathcal{S}_{{\bf h}, {\bf g}} T({\bf g}) e^{-2 \pi i {\bf g} \cdot {\bf R}} \right|^2 \;.
\end{equation}

Together, Eqs. (\ref{eq:a_matrix}), (\ref{eq:a_to_s}) and (\ref{eq:4d_sm}) solve the forward problem of simulating 4D STEM intensities from a given specimen projected potential, as shown schematically in the top row of Fig. \ref{fig:intro_schematic}. One approach to structure determination is to solve the forward problem for a model structure: if the simulated 4D STEM data is in close agreement with the experimental data then that increases confidence in the model structure.

\subsection{\Large The inverse problem} \label{sec:intro_inversion}

An alternative approach to structure determination is to solve the inverse problem: start with measured 4D STEM intensities and ``undo'' the forward operations to solve for the sample potential. Inverse multislice ptychography does this within the multislice formulation of the forward problem, separating the bulk specimen into a series of transmission and propagation steps and solving for the sample potential by using gradient descent to back-propagate the errors between forwards-simulated 4D STEM data and experiment \citep{maiden2012ptychographic,chen2021electron,sha2022deep,diederichs2024exact,ribet2024uncovering,zhu2025insights,clark2025electron}. \citet{ooe2025} similarly solve directly for the Fourier coefficients of potential from $I_{\rm 4D}$ within a scattering matrix formulation of the forward problem. However, here we instead follow the approach of \citet{brown2018} and \citet{sadri2023determining} which determines {\sm} as an intermediate step. The decomposition into the separate steps of determining {\sm} and then {\am} reduces the computational complexity of those individual steps (with parameters like thickness and mistilt only needing to be considered in the step of determining {\am} from {\sm}), and provides additional insight into, and ways to gauge the success of, the reconstruction process. Other potential advantages were discussed in the introduction.

The first step, starting with measured 4D STEM intensities and using Eq. (\ref{eq:4d_sm}) to solve for {\sm}, is a phase retrieval problem. \citet{brown2018} and \citet{pelz2021phase} used through-focal series phase retrieval, which requires a series of 4D STEM intensities taken across multiple focal planes. \citet{pelz2021phase} and \citet{schloz2024improved} showed that the additional overdetermination provided by data from multiple defocus values makes the phase retrieval more robust, including in the presence of partial coherence, though it introduces complications like accurately aligning the data from different defocus values and managing the total dose to avoid sample damage. \cite{findlay2021scattering} proposed that a single defocus value should suffice: with the aperture function in Eq. (\ref{eq:4d_sm}) acting as a compact support, a coupled phase retrieval algorithm of hybrid input-output and Gerchberg-Saxton error reduction can be applied to each virtual STEM image resulting from each pixel on the detector. The $T({\bf g})$ factor in Eq. (\ref{eq:4d_sm}) means that columns of {\sm} for indices ${\bf g}$ for which $|{\bf g}|>q_{\rm max}$ (where $q_{\rm max}$ denotes the reciprocal space cut-off due to the probe-forming aperture), indicated by the shaded vertical bands depicted on {\sm} in the inverse procedure in Fig. \ref{fig:intro_schematic}, cannot be retrieved. This has been called the truncation problem \citep{allen2001structure,findlay2021scattering}. Further, since the virtual STEM image from each detector pixel ${\bf h}$ is subject to an independent phase retrieval problem and since (inline holography type) phase retrieval can only ever retrieve the phase up to an arbitrary additive constant, there is potentially an unknown relative phase factor, $\phi_{{\bf h}}$, between the rows of the retrieved {\sm}. Finally, the lens aberrations may not be known. Consequently, the relation between the retrieved scattering matrix $\mathcal{S}^R$ and the true scattering matrix $\mathcal{S}$ is given by
\begin{equation}\label{eq:sm_recon}
    \mathcal{S}_{{\bf h}, {\bf g}}^R = e^{i \phi({\bf h})} \mathcal{S}_{{\bf h}, {\bf g}} T({\bf g}) \;.
\end{equation}
If the aberrations are known or negligible, the phase ambiguity within the bright field region can be resolved via a so-called antidiagonal symmetry property \citep{allen1999,allen2000,brown2018}, which can further be used to provide a good estimate for the sample thickness \citep{findlay2021scattering}. In later sections we explore the consequences of unknown lens aberrations being present and discuss how, when solving for the structure, we can use information in the dark field region. 

The second step, starting with a reconstructed {\sm} and using Eq. (\ref{eq:a_to_s}) to solve for {\am} and so for the projected potential, can be approached in various ways. \citet{allen1999,allen2000} and \citet{brown2018} showed an inversion to determine {\am}, and thus the Fourier coefficients of potential, that requires solving only linear equations. However, because the truncation problem can introduce appreciable artefacts within that approach, we instead use the optimisation-based approach proposed by \citet{sadri2023determining}. Using analytical derivatives, this approach consists of applying nonlinear conjugate gradient descent to find the Fourier coefficients of potential, sample thickness, probe defocus and unknown phase factors $\phi_{\bf h}$ that minimise an error metric between the prediction of Eq. (\ref{eq:a_to_s}) and the retrieved {\sm} in Eq. (\ref{eq:sm_recon}). \citet{sadri2023determining} showed that the constraint of Eq. (\ref{eq:a_to_s}) relating {\am} to {\sm} was more effective than that of Eq. (\ref{eq:4d_sm}) relating {\sm} to $I_{\rm 4D}$, and consequently, especially in the presence of significant noise, the reconstruction could be improved iteratively by using the determined {\am} to simulate {\sm} and using that to re-seed the phase retrievals, as depicted by the orange arrow inversion loop in Fig. \ref{fig:intro_schematic}. Stopping criteria were developed based on the Pearson correlation between the synthesised 4D STEM intensities across the various iterations and the experimental 4D STEM intensities.

Stagnation in local minima is a well-known phenomenon in multi-dimensional optimisation. Particularly from a poorly determined {\sm}, if we seek to optimise for all Fourier coefficients from an outset initialisation of zero potential the algorithm plays off different Fourier coefficients against each other, compensating under-converged low-order Fourier coefficients by significantly overestimating (the magnitude of) high-order Fourier coefficients. \citet{sadri2023determining} mitigated this by limiting the size of {\am}, starting small but increasing across subsequent iterations. We have since found it more effective to use a fixed, large matrix order throughout but to apply an apodisation function to the relative learning rates of the Fourier coefficients $U_{\bf g}$ based on $|{\bf g}|$. Using a top-hat-like function that exponentially decays to zero beyond a given frequency, we constrain the algorithm to only update those Fourier coefficients that fall inside the range of the apodisation function. This learning rate envelope is restricted to approximately 
$q_{\rm max}/2$ (i.e. half the reciprocal space cut-off due to the probe-forming aperture) for the first three iterations of the algorithm, which helps the low-order Fourier coefficients converge to reasonable values, before expanding out the spatial frequencies to some desired limit over the next three iterations. The apodisation function also significantly stabilises the convergence behaviour of reconstructions on simulated 4D STEM data with convergence semiangles as low as 12 mrad, as shown in a later section.

\subsection{\Large Thermal scattering}

\begin{figure}[htp]
    \centering
    \includegraphics[width = \linewidth]{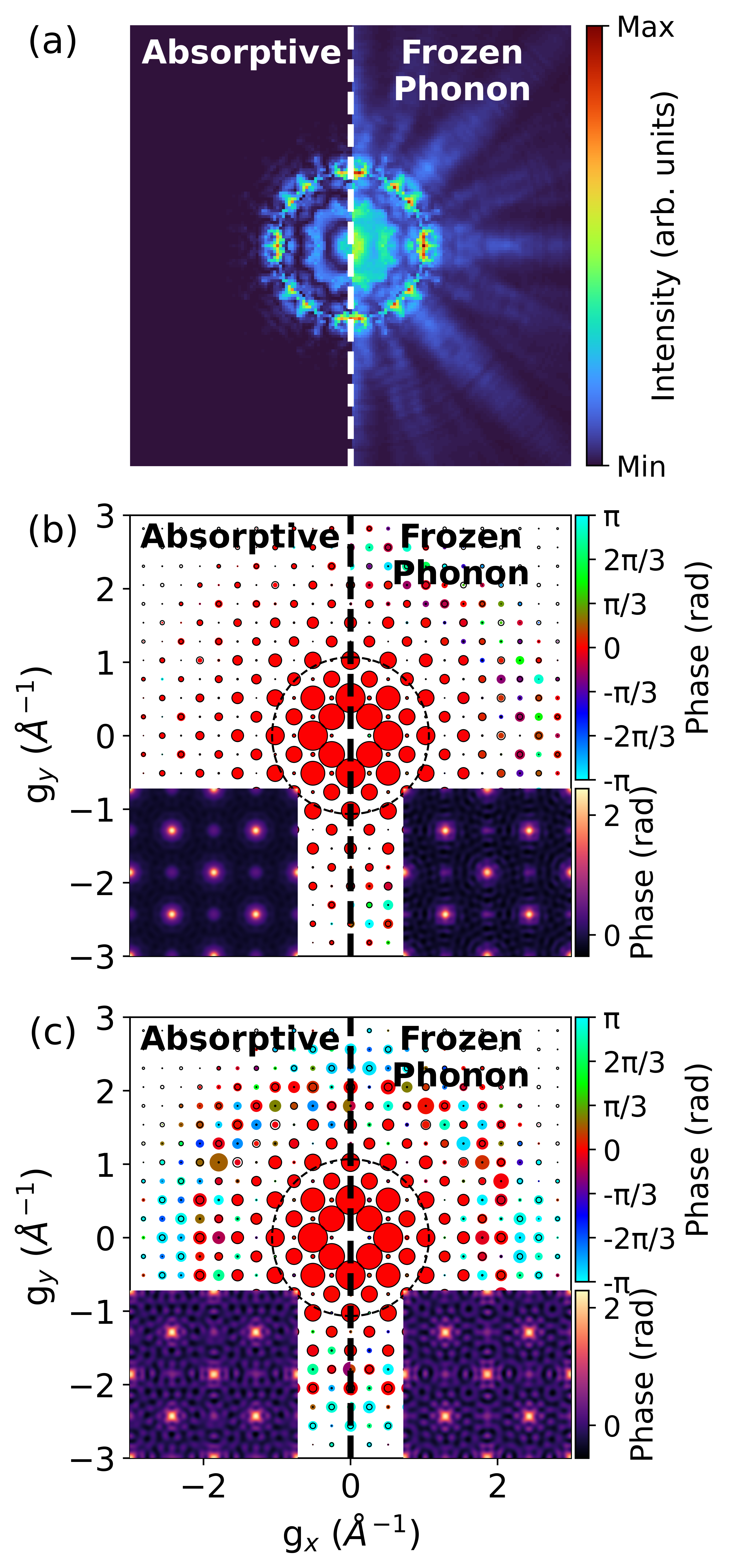}
    \caption{(a) Comparison of simulated CBED patterns between the absorptive model (left half) and frozen phonon model (right half) from a Sr column. The patterns are shown on a common colour scale, though the colours enhance the visibility of modest variations in intensity.
    (b) Comparison of the reconstructed electrostatic potential in both a reciprocal space visualisation (the scatter plot depicts the Fourier coefficients of potential, with accurate reconstructions appearing as red disks exactly filling black circles) and the corresponding real space visualisation (inset images). The dashed circle indicates the extent of the probe-forming aperture. A $6\times6$ supercell was used in the forward problem, but in the inversion problem the diffraction patterns were sub-sampled at every third pixel in each linear direction. (c) As (b), but using $3\times3$ binning in reciprocal space rather than subsampling.}
    \label{fig:frozen_phonon_vs_abs}
\end{figure}

\citet{diederichs2024exact} and \citet{herdegen2024thermal} recently showed through simulation and experiment that the model mismatch between including and neglecting the positive contribution of thermally scattered electrons to 4D STEM intensities could be a source of error in structure determination via ptychography. The scattering matrix formulation presented here partially accounts for thermal scattering via the absorptive model: the $U'_{\bf g}$ terms in Eq. (\ref{eq:a_matrix}) remove that portion of the electron density that undergoes inelastic thermal scattering from the elastic wavefield, but do not account for the positive contribution of thermally scattered electrons to the 4D STEM intensity. Let us check what impact this has on structure determination.

The 4D STEM simulations in this work were generated using the open-source \emph{py\_multislice} package \citep{pyms}. All simulations assume a SrTiO$_3$ sample oriented along the [001] zone axis and an accelerating voltage of 300 kV (matching the experimental data shown later). To examine the effect of model mismatch arising from incongruent models of thermal scattering, two calculations were performed, one using the frozen phonon model, which includes the positive contribution from thermally scattered electrons to the 4D STEM intensities, and another using the absorptive model, which does not. We assume a 200~{\AA} thick sample, a 21 mrad convergence semiangle and an aberration-free probe. Figure \ref{fig:frozen_phonon_vs_abs}(a) compares the resultant convergent-beam electron diffraction (CBED) pattern with the probe directly atop the Sr column. The positive contribution of thermally scattered electrons in the frozen phonon model is most evident in the Kikuchi lines extending symmetrically outwards from the centre of the pattern into the dark field, but modest intensity differences are also present in the bright-field disk. Since the Sr column is the heaviest scatterer present, the average difference across all probe positions will be lower than that in Fig. \ref{fig:frozen_phonon_vs_abs}(a). 

Figure \ref{fig:frozen_phonon_vs_abs}(b) and (c) compare the results of structure determination between the absorptive and frozen phonon simulated 4D STEM data.  The forwards calculations use a $6\times6$ supercell, but prior to the inversion problem Fig. \ref{fig:frozen_phonon_vs_abs}(b) reduced the diffraction patterns by sub-sampling at every third pixel in each (reciprocal) linear direction while Fig. \ref{fig:frozen_phonon_vs_abs}(c) reduced the diffraction patterns by $3\times3$ binning. These simulations neglect effects like shot noise or spatial incoherence (considered later), so in Fig. \ref{fig:frozen_phonon_vs_abs}(b) the main source of model mismatch is the difference in how thermal scattering is modelled, while in Fig. \ref{fig:frozen_phonon_vs_abs}(c) model mismatch derives from both thermal scattering and binning. These figures depict both a Fourier space representation of the reconstructed potential (the plot) and a real space representation (inset image). In the Fourier reconstruction, the small black circles indicate the reference (input) values of the Fourier coefficients, centred on the location of the spatial frequency they represent and with area proportional to the Fourier coefficient magnitude. The coloured disks indicate the reconstructed Fourier coefficients, with area proportional to the Fourier coefficient magnitude and colour showing the phase difference between the reconstructed and reference potential. An accurately reconstructed Fourier coefficient will thus appear as a red disk exactly filling a black circle. The dashed circle in the Fourier space visualisation is used to indicate the extent of the probe-forming aperture.

Figure \ref{fig:frozen_phonon_vs_abs}(b) shows that there is little difference in the fidelity of the reconstructions from the two different 4D STEM intensity datasets. Indeed, the only perceptible differences are in the higher-order Fourier coefficients. Figure \ref{fig:frozen_phonon_vs_abs}(c) shows similarly good fidelity for low-order Fourier coefficients, but increased error in high-order Fourier coefficients, which we attribute to binning effectively reducing the coherence in the phase retrieval step. In later sections we will see that introducing experimental effects like noise and spatial incoherence introduce further errors in those Fourier coefficients. Given these findings, any appreciable limitations on subsequent reconstructions should not be attributed to model mismatch due to thermal scattering. Nevertheless, all other simulated 4D STEM datasets in this work are simulated using the frozen phonon model, being more accurate and avoiding concerns around using the same model for the forward and inverse problem \citep{kaipio2007statistical}. Note that the degree of thermal scattering in our examples is likely lower than that of \citet{diederichs2024exact}, who considered a lower accelerating voltage and a sample including atoms with higher atomic number. Our findings are consistent with the results of \citet{herdegen2024thermal}, with our implicit ``regularisation'' of built-in periodicity and the use of a Fourier basis with apodisation being reminiscent of their sparse frequency regularisation.

Figure \ref{fig:frozen_phonon_vs_abs}(b) and (c) show the reconstructed elastic component of the sample potential (i.e. $U_{\bf g}$ from Eq. (\ref{eq:a_matrix})). We also reconstruct the absorptive potential ($U'_{\bf g}$ in Eq. (\ref{eq:a_matrix})), but, being about an order of magnitude weaker than the elastic potential, its reconstruction is less reliable than that of the elastic potential, particularly under the influence of spatial incoherence and shot noise considered in subsequent sections, and so is not shown.

\section{\LARGE Improving the reconstruction algorithm}\label{sec:innovations}

\subsection{\Large Spatial incoherence}\label{sec:spatinc}

The 4D STEM intensity expression of Eq. (\ref{eq:4d_sm}) assumes perfect coherence. However, in practice various microscope instabilities limit the degree of coherence in the recorded data. Partial spatial coherence can be incorporated by convolving Eq. (\ref{eq:4d_sm}) with an effective source distribution\footnote{The effective source distribution is often denoted $S$, but we adopt a different notation to avoid confusion with the scattering matrix {\sm}.} $\Lambda({\bf R})$ that encompasses not just the finite size of the electron emitter but also instabilities in the microscope that blur the signal \citep{nellist1994beyond}:
\begin{align}\label{eq:spatialinc}
    I_{\rm 4D}({\bf h}, {\bf R}) &= \left| \sum_{{\bf g}} \mathcal{S}_{{\bf h}, {\bf g}} T({\bf g}) e^{-2 \pi i {\bf g} \cdot {\bf R}} \right|^2 \otimes_{{\bf R}} \Lambda({\bf R}) \;,
\end{align}
where $\otimes_{{\bf R}}$ denotes convolution with respect to ${\bf R}$. Partial spatial coherence acts something like a low-pass filter, blurring sharp coherent interference features in individual CBED patterns by averaging the CBED patterns over a range of probe positions. Demagnifying the source can mitigate these effects by reducing the contribution of the emitter size/shape but does not eliminate the contribution from instabilities.

In their {\sm}-matrix structure determination, \citet{brown2018} corrected for spatial incoherence  by deconvolving an assumed effective source distribution from each detector pixel STEM image, essentially attempting to invert Eq. (\ref{eq:spatialinc}). The algorithm development and predictions of \citet{findlay2021scattering} and \citet{sadri2023determining} neglected spatial incoherence on the basis that it could be tackled in that way, despite deconvolution being inherently unstable. However, \citet{clark2011simultaneous} proposed a more stable approach that can be readily adapted within the hybrid input-ouput phase retrieval algorithm we use to reconstruct {\sm} from 4D STEM. It consists of modifying the amplitude update step to be\footnote{The wavefield here is the quantity within the modulus-square in Eq. (\ref{eq:4d_sm}). Effecting phase retrieval by using the probe-forming aperture for compact support, Eq. (\ref{eq:clarkpeele}) regards ${\bf h}$ parameterically and ${\bf R}$ as a coordinate. No such wavefield is present in the actual STEM experiment, but the mathematical structure is nevertheless one to which well-established phase retrieval techniques apply.}
\begin{equation}\label{eq:clarkpeele}
    \psi^{j+1}({\bf h}, {\bf R}) = \psi^{j}({\bf h}, {\bf R}) \sqrt{ \frac{ I_{\rm 4D}({\bf h}, {\bf R}) }{ |\psi^{j}({\bf h}, {\bf R})|^2 \otimes_{\bf R} \Lambda({\bf R}) } } \;.
\end{equation}
This approach assumes the coherence function $\Lambda({\bf R})$ to be fairly well known. The extent of the effective source for a given experiment is often not something we know \emph{a priori}. Various strategies exist for measuring the effective source size/distribution \citep{maunders2011practical,dwyer2012sub}, though doing so within the optical configuration of a standard STEM experiment requires comparison against simulation using a known structure. Keeping within the bounds of the {\am} matrix reconstruction workflow outlined in Fig. \ref{fig:intro_schematic}, we may instead estimate the effective source size by solving the inverse scattering problem for a sequence of effective source distributions refined for successively better fit to the experimental data, a grid search strategy which effectively minimises the difference between measured and simulated 4D STEM intensities, as will be shown later when we consider the experimental data.

\begin{figure*}[ht]
    \centering
    \includegraphics[width = \textwidth]{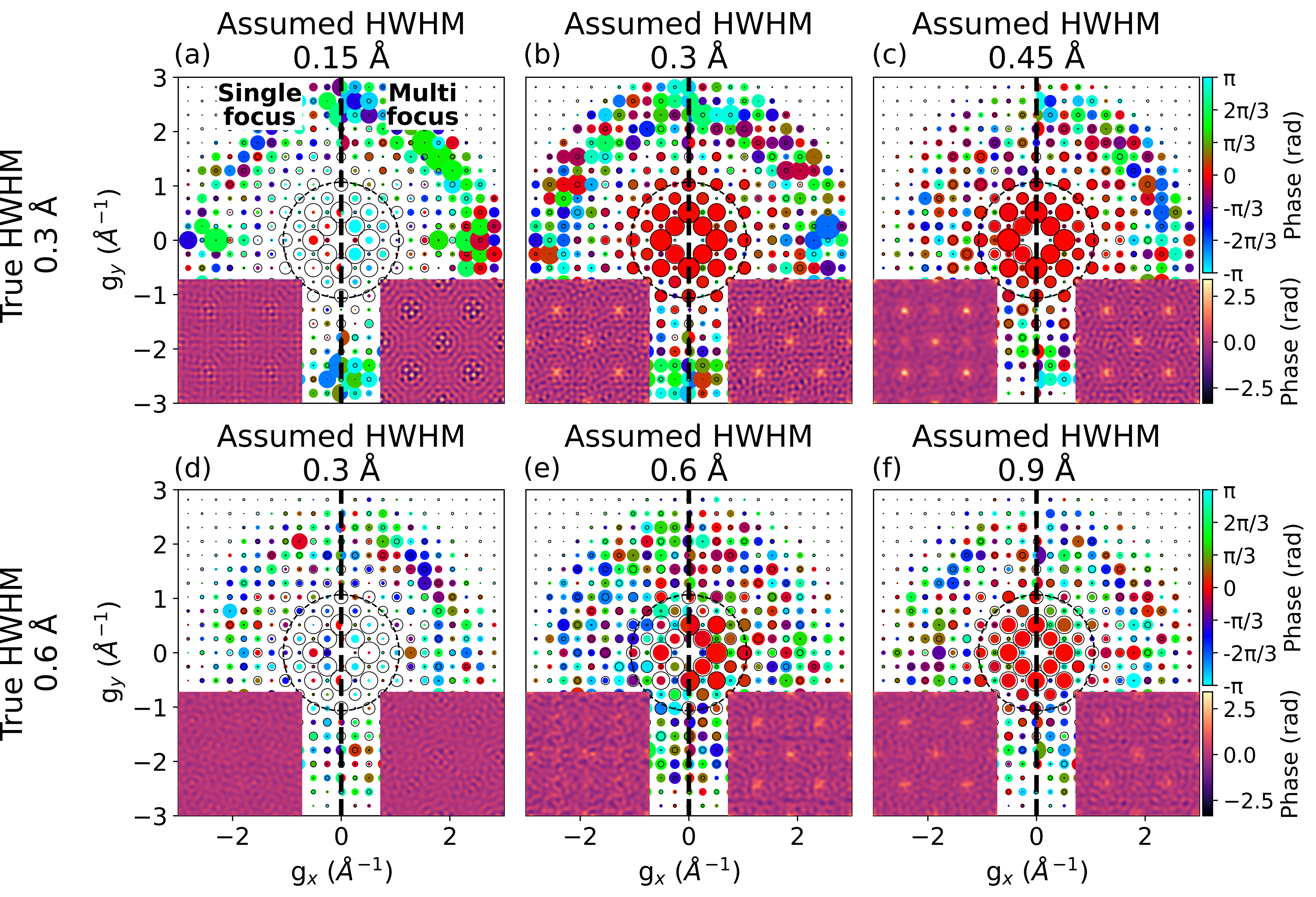}
    \caption{From 4D STEM data incorporating a Gaussian effective source distribution with a small (0.3~{\AA}; (a-c)) and large (0.6~{\AA}; (d-f)) HWHM, we compare structure determination results when the HWHM is underestimated (by 50\%; (a,d)), accurate (b,e) or overestimated (by 50\%; (c,f)). The left pane of each image uses data at a single defocus value (0~{\AA}), while the right pane uses data from multiple defocus values (0~{\AA} and 50~{\AA} over focus).}
    \label{fig:spatinc_sm_am}
\end{figure*}

To explore this approach, we simulated 4D STEM data incorporating spatial incoherence (via Eq. (\ref{eq:spatialinc})) for a Gaussian effective source with a half-width-at-half-maximum (HWHM) of 0.3~{\AA} and 0.6~{\AA}, and an incident dose level of 1000 C/cm$^2$ ($6.2\times10^5$ e$^{-}$/{\AA}$^2$). Reconstructions were performed for three different scenarios: underestimating the effective source HWHM by 50\%, knowing the effective source distribution exactly, and overestimating the true HWHM by 50\%. The results of these reconstructions are displayed in Fig. \ref{fig:spatinc_sm_am} via the reciprocal space visualisation of the Fourier coefficients of the potential, with the real space visualisation inset. We also compare reconstructions from 4D STEM data at a single defocus (0~{\AA}) with that from data at two defoci ($0$~{\AA} and $50$~{\AA})\footnote{We adopt the defocus sign convention where positive values indicate overfocus, and negative values indicate underfocus.} in the left and right panes, respectively, of each image in Fig. \ref{fig:spatinc_sm_am}, since additional phase diversity may mitigate the effects of partial spatial coherence \citep{pelz2021phase,schloz2024improved}. We adopt dose fractionation in the case of two defoci to ensure the total dose incident on the sample remains constant between the two datasets, with each defocus therefore simulated with an incident dose level of 500 C/cm$^2$ (3.1$\times 10^5$ e/{\AA}$^2$). 

Figures \ref{fig:spatinc_sm_am} (a,d) show that the reconstructions fail when we underestimate the effective source size by 50\%, displaying both erroneous magnitude and phase across the full range of spatial frequencies. By contrast, for the 0.3~{\AA} HWHM case both accurate (Fig. \ref{fig:spatinc_sm_am} (b)) and overestimated (Fig. \ref{fig:spatinc_sm_am} (c)) HWHM values lead to good reconstruction of the low-order Fourier coefficients (those near and within the bright-field disk), shown as red discs closely filling the black outline. The higher order Fourier coefficients and those with very small amplitudes are less reliable, varying significantly in both phase and amplitude. Although extending the apodisation function to include these higher order frequencies increases the Pearson correlation against the measured 4D STEM intensities, this appears to be overfitting to noise rather than improving the structural information. The errors in the high-order Fourier coefficients manifest on the inset real space reconstructed potentials as high frequency oscillations. These oscillations obscure the oxygen columns, but this is arguably a limitation of the real space visualisation, which could be ameliorated via low pass filtering. As the Fourier representation makes clear, there is a significant difference in reliability of the low-order Fourier coefficients compared to the high-order Fourier coefficients.

The reconstructions for the 0.6~{\AA} HWHM case (Fig. \ref{fig:spatinc_sm_am} (e,f)) are worse than those for the 0.3~{\AA} HWHM case (Fig. \ref{fig:spatinc_sm_am}(b,c)). Indeed, for a single defocus the 0.6~{\AA} HWHM reconstruction fails even when the HWHM is accurately known. Less expected is that for the 0.6~{\AA} HWHM case, overestimating the HWHM (Fig. \ref{fig:spatinc_sm_am} (f)) results in a better reconstruction than accurately estimating the HWHM value (Fig. \ref{fig:spatinc_sm_am} (e)). It is hard to gauge the generality of the result without exhaustive parameter space searching, but Fig. \ref{fig:spatinc_sm_am} suggests that in this approach to accounting for spatial incoherence it is better to overestimate the width of the effective source distribution than to underestimate it.

The reconstructions for the 0.6~{\AA} HWHM case with the HWHM accurately known fails for a single defocus, while yielding a somewhat reasonable reconstruction for the largest, low-order Fourier coefficients when data from two defocus values is used (Fig. \ref{fig:spatinc_sm_am}(e)). However, in the other cases the added phase diversity and data redundancy of having data from two defocus values is insufficient to appreciably improve the reconstruction fidelity, or to transform a failed reconstruction into a successful one: the gains from using two defocus values over a single defocus value are marginal (the potentials reconstructed using through-focal data have smaller errors, by almost a factor of 2 in the case of gradient descent cost function for {\sm}, but this is not manifesting as clear improvement in the accuracy of the Fourier coefficients). Previous work suggests this might improve with data from several defocus values \citep{pelz2021phase,schloz2024improved}, but only at the expense of increased difficulties in the data acquisition and at increased computational cost. We rather conclude from Fig. \ref{fig:spatinc_sm_am} that a single defocus value should suffice provided spatial incoherence is not too large and that we account for it by erring on the side of overestimation when the effective source distribution is not accurately known.

It is worth comparing the results in Figs. \ref{fig:spatinc_sm_am}(b,c,e,f), where higher-order Fourier coefficients are erroneous in both amplitude and phase, to Fig. \ref{fig:frozen_phonon_vs_abs}(b) with perfect coherence but no noise, where the Fourier coefficients across the full range of allowed spatial frequencies are accurately reconstructed. Even when the spatial incoherence function $\Lambda({\bf R})$ is known precisely, the averaging across CBED patterns effected by spatial incoherence results in an irreversible loss of high frequency coherent information (as distinct from noise, where the variability of noise realisations across different CBED patterns might still allow reasonable reconstruction of some high frequency information). Spatial incoherence will also likely limit how thick a sample this approach can reliably handle. \citet{brown2018} showed through simulation the need for finer reciprocal space sampling as the sample thickness increases. Defocus and other probe aberrations also often introduce fine features. Since spatial incoherence smooths out fine features in the CBED pattern, the model mismatch arising from spatial incoherence is likely larger for thicker samples or data at larger defocus values.

\subsection{\Large Thickness estimation in the presence of unknown probe aberrations}
\label{sec:thicknessest}

The {\am}-matrix (Eq. (\ref{eq:a_matrix})), which assumes the projected potential approximation and thus a crystal with a small (sub-nanometer) repeat distance along the optical axis, possesses the so-called antidiagonal symmetry property\footnote{The terminology ``antidiagonal symmetry'' follows from ordering the matrix elements such that ${\bf g}={\bf 0}$ is the central index, and vectors ${\bf g}$ and $-{\bf g}$ are symmetrically spaced indices about that centre. However, the identity written in the subscript form $\mathcal{A}_{{\bf h},{\bf g}} = \mathcal{A}_{-{\bf g}, -{\bf h}}$ is true regardless of the ordering of matrix elements chosen.} $\mathcal{A}_{{\bf h},{\bf g}} = \mathcal{A}_{-{\bf g}, -{\bf h}}$. The scattering matrix for a single crystal (Eq. (\ref{eq:a_to_s})) then shares this property:
\begin{equation}\label{eq:sm_antidiag}
    \mathcal{S}_{{\bf h}, {\bf g}} = \mathcal{S}_{-{\bf g}, -{\bf h}} \;.
\end{equation}

\citet{allen1999,allen2000} showed that this property could be used to resolve the unknown phase factors that arise when determining {\sm} via phase retrieval, here the $\phi_{\bf h}$ in Eq. (\ref{eq:sm_recon}) relating the retrieved matrix $\mathcal{S}^R$ to the true matrix $\mathcal{S}$.\footnote{In the conventional TEM context of \citet{allen1999,allen2000} it is the columns of {\sm} that need to be correctly phased after phase retrieval. In the STEM context, it is the rows that need to be correctly phased (Eq. (\ref{eq:sm_recon})).} \citet{findlay2021scattering} showed through simulation that determining the $\phi_{\bf h}$ by enforcing antidiagonal symmetry should provide a thickness estimate accurate to within a couple of nanometers. That work assumed an aberration-free probe, though would also apply for an aberrated probe if the aberrations were accurately known. However, probe aberrations, especially defocus, are seldom accurately known in practice. Let us explore what constraints antidiagonal symmetry provides when unknown aberrations are present.

The phase retrievals for reconstructing {\sm} are initialised by setting $\mathcal{S}_{{\bf h}, {\bf g}}^R = \delta_{{\bf h}, {\bf g}}$. Simulations show that this uniform (zero) phase along the diagonal of {\sm} largely persists through to the reconstruction \citep{findlay2021scattering}. It then follows from Eq. (\ref{eq:sm_recon}) that
\begin{align}\label{eq:phih}
    0 &\approx \arg\left[ \mathcal{S}_{{\bf h}, {\bf h}}^R \right] = \arg\left[ e^{i \phi_{\bf h}} \mathcal{S}_{{\bf h}, {\bf h}} T({\bf h}) \right] \nonumber \\
    0 &\approx \phi_{\bf h} + \arg\left[ \mathcal{S}_{{\bf h}, {\bf h}} \right] + \arg\left[ T({\bf h}) \right] \nonumber \\
    \phi_{\bf h} &\approx \frac{\pi t h^2}{K} + 2 \pi K \chi({\bf h}) \;,
\end{align}
where the final line follows because the phases of the diagonal elements of {\sm} are dominated by free-space propagation, i.e. $\arg\left[ \mathcal{S}_{{\bf h}, {\bf h}} \right] \approx - \frac{\pi t h^2}{K}$ \citep{findlay2021scattering}. We stress that we are assuming the aberration function $\chi({\bf h})$ contained in $T({\bf h})$ represents the difference between what (if any) aberration parameters are assumed to be present and those that are actually present.

Substituting Eq. (\ref{eq:phih}) into Eq. (\ref{eq:sm_recon}) gives the following relation between the retrieved and true {\sm}:
\begin{equation}\label{eq:sm_recon_explicit}
    \mathcal{S}_{{\bf h}, {\bf g}}^R \approx e^{i \pi t h^2 / K} T^*({\bf h}) \mathcal{S}_{{\bf h}, {\bf g}} T({\bf g}) \;,
\end{equation}
where we have written $T^*({\bf h})$ instead of $\exp\left[i \frac{2\pi}{\lambda} \chi({\bf h})\right]$ both for notational simplicity and to emphasise that the result only applies for ${\bf h}$ within the bright field region. (We return in the next section to how to handle the dark field region, which antidiagonal symmetry does not constrain.) Eq. (\ref{eq:sm_recon_explicit}) shows that the retrieved {\sm} does not possess antidiagonal symmetry. We can try to impose antidiagonal symmetry via the degree of freedom of multiplying each row by a phase factor $e^{i\psi_{\bf h}}$ (which, as per Eq. (\ref{eq:4d_sm}), would not change the 4D STEM intensities and so is an equally valid solution of the phase retrieval step) by solving
\begin{align} \label{eq:phasing_argmin}
    \{\psi_{\bf h}\} = \underset{\{\psi_{\bf h}\}}{\mathrm{argmin}} \left|e^{i \psi_{\bf h}} \mathcal{S}_{{\bf h}, {\bf g}}^R - e^{i \psi_{-{\bf g}}} \mathcal{S}_{-{\bf g}, -{\bf h}}^R \right|^2 \;.
\end{align}
If Eq. (\ref{eq:sm_recon_explicit}) held exactly then the analytic solution would be
\begin{equation}\label{eq:phase_factor_general}
    e^{i \psi_{\bf h}} \approx e^{-i \pi \lambda t h^2} T({\bf h}) T(-{\bf h}) \;,
\end{equation}
as readily shown by noting the resultant phased {\sm}
\begin{align}\label{eq:sm_recon_phased}
    \mathcal{S}_{{\bf h}, {\bf g}}^P = e^{i \psi_{\bf h}} \mathcal{S}_{{\bf h}, {\bf g}}^R = T(-{\bf h}) \mathcal{S}_{{\bf h}, {\bf g}} T({\bf g})
\end{align}
has antidiagonal symmetry. Eq. (\ref{eq:sm_recon_phased}) shows that enforcing antidiagonal symmetry does not eliminate the effect of unknown lens aberrations, which therefore need to be included in the optimisation solving for the structure \citep{sadri2023determining}. However, solving for the structure is aided by having a good starting guess for the aberrations, and the benefit of antidiagonal symmetry comes from what we can deduce from Eq. (\ref{eq:phase_factor_general}) having determined the $\psi_{\bf h}$ via Eq. (\ref{eq:phasing_argmin}).

The odd-order aberrations cancel out in Eq. (\ref{eq:phase_factor_general}) and so are not constrained. Fortunately, the effect of odd-order aberrations are more apparent during an atomic resolution imaging experiment (leading to non-round distortions of atomic columns), and so may be more readily corrected before data acquisition. The dominant even-order aberration is defocus, $\Delta f$, which can indeed be hard to estimate in practice. Assuming defocus to be the only significant (unknown) aberration, Eq. (\ref{eq:phase_factor_general}) yields
\begin{equation}\label{eq:phase_factor_specific}
    e^{i \psi_{\bf h}} \approx e^{-i \pi \lambda (t + 2\Delta f) h^2} \;.
\end{equation}
If $\Delta f=0$ {\AA}, this reduces to the result of \citet{findlay2021scattering} that fitting a parabola to $\psi_{\bf h}$ provides an estimate of the thickness. With some unknown, non-zero defocus, fitting a parabola to $\psi_{\bf h}$ will instead provide an estimate for $t + 2 \Delta f$. This is not as good as an estimate for either (or both) individually, but is still a useful constraint: if the strategy for estimating these value involves trying multiple initialisations, either via simple grid search \citep{sadri2023determining} or Bayesian optimisation \citep{cao2022automatic}, then having an approximate constraint linking thickness and defocus reduces the sample space over which to search.

To explore the reliability of estimating $t+2\Delta f$, we again assume a 21 mrad probe-forming aperture semiangle and simulate 4D STEM intensities from SrTiO$_3$ [001] via the frozen phonon model. Spatial incoherence is incorporated assuming a Gaussian effective source with 0.3~{\AA} HWHM that we presume to know exactly when reconstructing {\sm}. Reconstructions are conducted across 100 different noise realisations at dose levels of 10 C/cm$^2$, 100 C/cm$^2$, and 1000 C/cm$^2$ (equivalent to 6.2$\times 10^3$ e$^-$/{\AA}$^2$, 6.2$\times 10^4$ e$^-$/{\AA}$^2$ and 6.2$\times 10^5$ e$^-$/{\AA}$^2$ respectively), and across a systematic range of thickness and defocus combinations. Figure \ref{fig:antidiag} shows box plots summarising the distribution of thickness estimates that result from the different noise realisations by applying the antidiagonal symmetry to estimate $t+2\Delta f$ (Eqs. (\ref{eq:phasing_argmin}) and (\ref{eq:phase_factor_specific})) and assuming we know defocus. (The data could be reinterpreted to estimate defocus if we assumed instead to know thickness.) The optimisation for finding thickness benefits from setting reasonable bounds within which the solution should fall, which here were set to $\pm 50\%$ of the true thickness.

\begin{figure*}[ht]
    \centering
    \includegraphics[width = \textwidth]{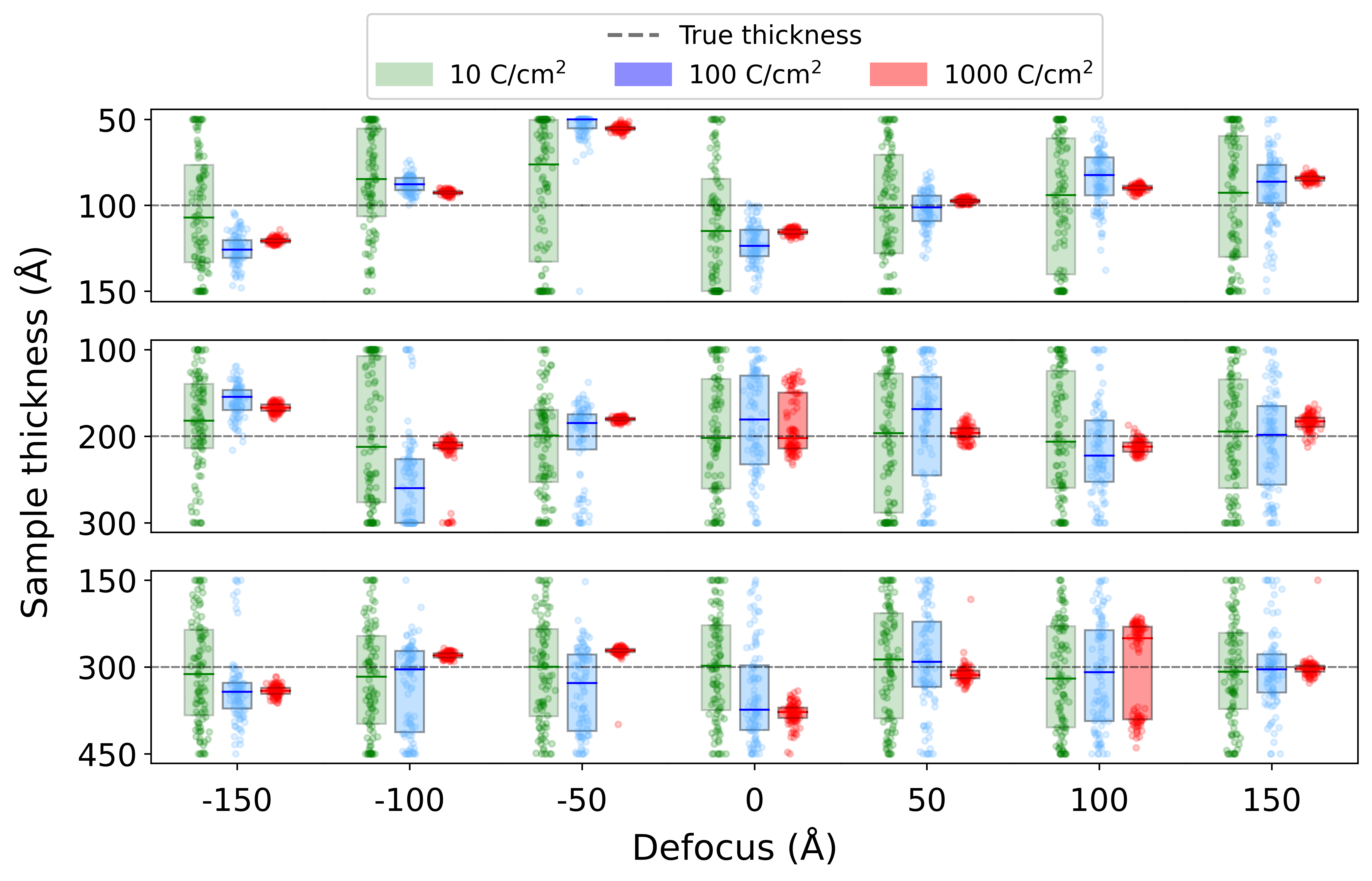}
    \caption{Box plots summarising the distribution of thickness values estimated by enforcing antidiagonal symmetry on $\mathcal{S}$-matrices reconstructed from 100 different noise realisations of simulated 4D STEM data, for a systematic range of thickness and defocus values and across three different dose levels. Antidiagonal symmetry provides an estimate for $t+2\Delta f$, which we here interpret as a thickness by assuming the defocus is known exactly. The dark coloured line in each box plot indicates the median value for each distribution. The box plots themselves show the interquartile range of the data.
    The dashed line indicates the true sample thickness.}
    \label{fig:antidiag}
\end{figure*}

Figure \ref{fig:antidiag} shows that for the highest dose the spread in estimated thicknesses is narrow (i.e. the estimate is precise) and in most cases falls within a couple of nanometers of the true thickness (i.e. the estimate is fairly accurate). The most accurate results here occur for 50~{\AA} overfocus, and in most cases overfocus is seen to produce more accurate estimates than underfocus. There are some anomalies. The 300~{\AA} thickness, 0~{\AA} defocus case and the 100~{\AA} thick, $-50$ {\AA} defocus cases show systematic thickness mis-estimates of several nanometers. The 300~{\AA} thickness, 100~{\AA} defocus estimates show a wide spread in the distribution of thicknesses, though in this case the box plot obscures a near bimodal distribution of thickness estimates. How and why phase retrievals fail to converge to the true solution is not easy to determine, but this behaviour is consistent with getting stuck in a local minimum. While we must be aware of this potential for anomalies, that accurate thickness estimates are predicted in most cases out to more than 100~{\AA} in defocus is encouraging, especially since defocus tends to introduce finer features in diffraction patterns that we might expect to be more affected by spatial incoherence.

Fairly consistently across thickness and defocus, the successively noisier, lower dose cases result in successively larger distributions of thickness estimates across the different noise realisations. While in some cases the mean value remains accurate to within a couple of nanometers, with only a single dataset (and so noise realisation) the estimate could be in error by several nanometers. \citet{sadri2023determining} predict that structure determination should be possible at lower doses, through the iterative process that includes re-seeding the step determining {\sm} (see Fig. \ref{fig:intro_schematic}). However, the reasoning of Eqs. (\ref{eq:phih})--(\ref{eq:phase_factor_specific}), being based on the original initialisation of the phase retrievals, only holds on the first retrieval of {\sm} from the 4D STEM intensities.\footnote{A variant on the antidiagonal symmetry argument can be applied to subsequent iterations for the residual error in thickness and aberrations, but doing so does not seem to offer much further diagnostic insight.} Therefore, one should be aware that the antidiagonal symmetry constraint is less effective at lower doses than it is at higher doses in limiting the search space when seeking promising initial estimates of thickness and defocus.

\subsection{\Large Dark-field reconstruction}
\label{sec:darkfield}

Values of $\mathcal{S}_{{\bf h},{\bf g}}$ for which ${\bf g}$ lies outside the probe-forming aperture do not contribute to, and so cannot be reconstructed from, the 4D STEM intensity (see Eq. (\ref{eq:4d_sm})). Figure \ref{fig:intro_schematic} indicates this by greyed out columns in {\sm} in the inverse problem. The sample-free initialisation of the phase retrieval problem (mathematically, $\mathcal{S}^R_{{\bf h},{\bf g}}=\delta_{{\bf h},{\bf g}}$) cannot initialise the phase retrieval of STEM images from detector pixels in the dark field region (Fig. \ref{fig:intro_schematic} indicates this by partially greyed out rows). Because of this, and because the \citet{allen1999,allen2000} algorithm for solving for {\am} given {\sm} involves an eigenproblem requiring a square matrix, the experimental structure reconstruction of \cite{brown2018} used only information from the bright field region. However, it has been recognised in ptychography \citep{jiang2018electron} and in the parallax reconstruction \citep{terzoudis2023resolution} that the lateral and depth resolution can be improved if the intensity in the dark field region is incorporated. Only considering the determination of {\sm}, \citet{terzoudis2023resolution} sought to overcome this via an iterative polynomial extrapolation from the bright field region in a representation of {\sm} that makes evident its continuity properties. However, extrapolation is unstable, and even in simulation that approach only looks promising for thin or weakly scattering samples. 

Solving for the structure enables an alternative approach by leveraging the iterative procedure by which we successively refine the projected potential of the sample. The initially reconstructed {\sm} uses only the bright-field region since we do not have information to phase the rows in the dark field region. But re-seeding the phase retrievals from a simulated {\sm} (calculated from the reconstructed {\am}) provides a phase relation between the dark field region and the bright field region. Thus, STEM images from the dark field region can then be included in the phase retrieval process, and the dark field rows included in the cost function for the nonlinear conjugate gradient descent step that refines {\sm}.

\begin{figure*}[ht]
    \centering
    \includegraphics[width=0.74\linewidth]{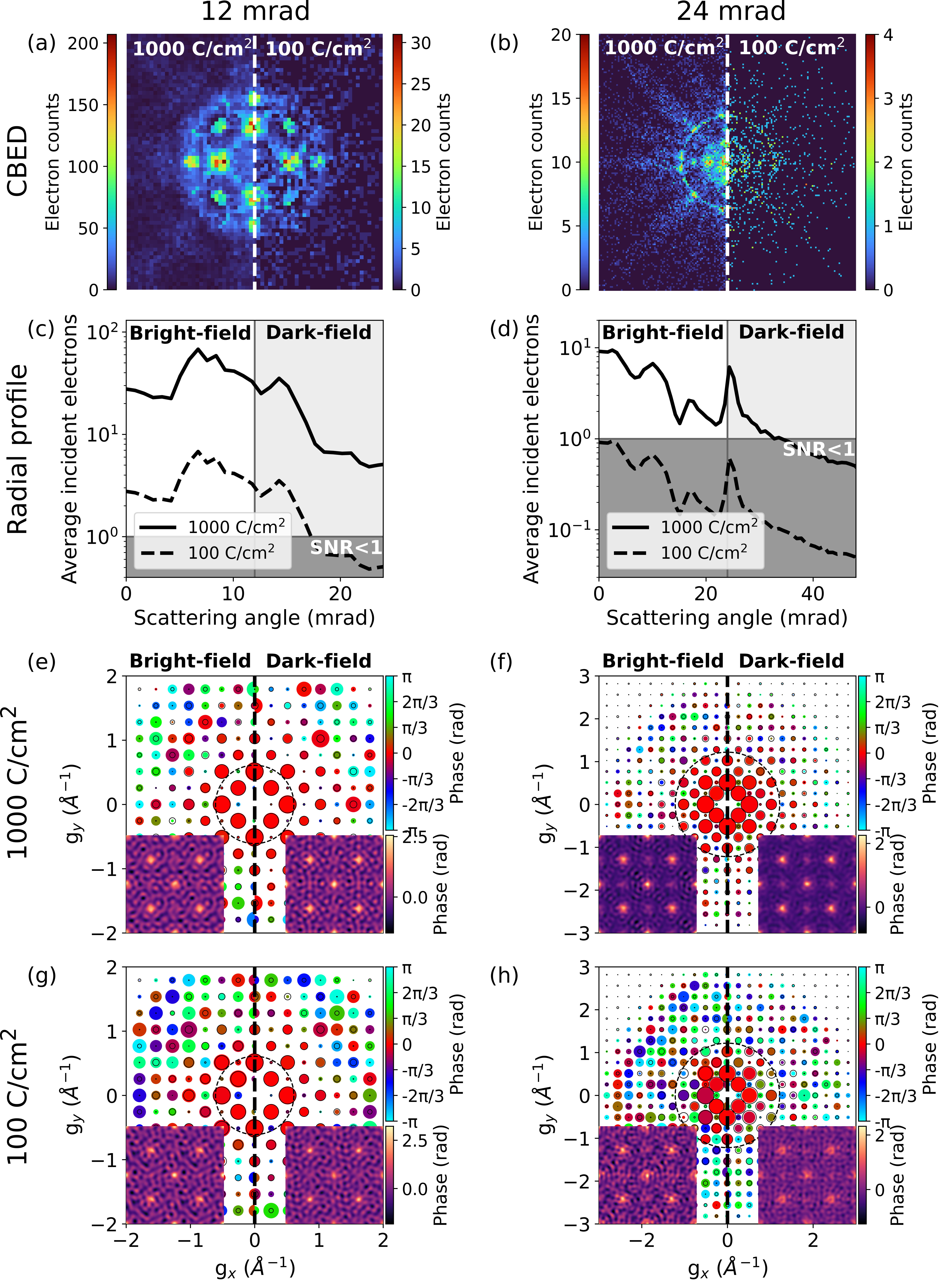}
    \caption{(a,b) Simulated CBED patterns at 300 keV with the probe atop a 200~{\AA} thick Sr column, with the left panes assuming a dose of 1000 C/cm$^2$ and the right panes a dose of 100 C/cm$^2$. (c,d) Radial averages of the simulated CBEDs from (a,b), plotting the average number of electrons per pixel for each dose level as a function of scattering angle out to twice the convergence semiangle. Dark grey shading indicates the region with less than 1 electron/pixel, implying a signal-to-noise (SNR) ratio less than one. (e,f) Visualisation of the potential reconstructed from the simulated 4D STEM data at 1000 C/cm$^2$, with the left panes based only on the bright field whereas the right panes include dark-field signal in the reconstructions. (g,h) As per (e,f), but with an incident dose of 100 C/cm$^2$. A 12 mrad convergence semiangle is used for (a,c,e,g) while a 24 mrad convergence semiangle is used for (b,d,f,h).
    }
    \label{fig:bf_vs_df_combined}
\end{figure*}

In perfect simulations we can extend far into the dark field region. In practice, larger matrices increases the computational cost but the greater limitation is shot noise. To illustrate this, Figs. \ref{fig:bf_vs_df_combined}(a,b) show simulated CBED patterns, assuming 12 mrad and 24 mrad convergence semiangles respectively, for a probe atop the Sr column in 200~{\AA} thick SrTiO$_3$ viewed along the [001] zone axis, with the left half assuming a dose of 1000 C/cm$^2$ (6.2$\times 10^5$ e$^-$/{\AA}$^2$) and the right half assuming a dose of 100 C/cm$^2$ (6.2$\times 10^4$ e$^-$/{\AA}$^2$). Figures \ref{fig:bf_vs_df_combined}(c,d) show the corresponding mean number of incident electrons per pixel as a function of scattering angle, which is seen to be typically lower in the dark-field than in the bright-field, tending to decrease with increasing scattering angle. Shot noise follows a Poisson distribution where the variance equals the mean. Consequently, the signal-to-noise ratio (SNR) drops below unity when the number of electrons per pixel drops below 1, indicated by the dark grey shading in Figs. \ref{fig:bf_vs_df_combined}(c,d). This occurs at a scattering angle of around 33 mrad for the 24 mrad convergence semiangle at the higher dose level, and across both the bright-field and dark-field signal at the lower dose level. The 12 mrad aperture only sees such a drop beyond 17 mrad at 100 C/cm$^2$. We should therefore expect noise to significantly affect the phase retrievals involving pixels in these regions.

To explore this, Figs. \ref{fig:bf_vs_df_combined} (e,f) show the results of structure retrievals from simulated noisy 4D STEM data, pertaining to 12 mrad and 24 mrad convergence semiangles respectively, assuming 1000 C/cm$^2$, and Figs. \ref{fig:bf_vs_df_combined} (g,h) show the equivalent figures assuming 100 C/cm$^2$. For each combination of incident dose and convergence angle we performed two reconstructions: one using bright-field information only (left half), and one leveraging dark-field information (right), keeping all other reconstruction parameters constant. Drawing on Figs. \ref{fig:bf_vs_df_combined}(c,d), when using dark-field information we consider only those beams out to a scattering angle 50\% beyond the bright-field disk radius, {i.e.} those extending out to 18 mrad for the 12 mrad convergence semiangle, and those extending out to 36 mrad for the 24 mrad convergence semiangle. All cases assume a Gaussian effective source with 0.3~{\AA} HWHM that we presume to know exactly.

Including the dark-field signal reduces errors in high-frequency Fourier coefficients, particularly in the 24 mrad case, extending the range of reasonably reconstructed Fourier coefficients out to higher scattering angles. The relative gains are less pronounced but still evident for the smaller (12 mrad) aperture size where the measured extent of ${\mathcal S}$ is particularly limited if dark field information is not used. The overall effect however is fairly modest, reflecting that binning, spatial incoherence and noise all limit reconstruction of high frequencies. 

Unsurprisingly, the reconstructions at higher dose are better than those at lower dose, but there is an interesting subtlety here. At 100 C/cm$^2$, the reconstructions for a 24 mrad convergence semiangle (Fig. \ref{fig:bf_vs_df_combined}(f)) are worse than those for 12 mrad (Fig. \ref{fig:bf_vs_df_combined}(g)). Here, not only the dose but also the diffraction pattern sampling is the same. Consequently, the narrower convergence angle concentrates more dose in a smaller number of pixels in the bright field and low angle dark field region, as shown in the radial profiles in Figs. \ref{fig:bf_vs_df_combined}(c,d): the average number of electrons per pixel throughout the bright field region is greater than 1 for the 12 mrad case but less than 1 for the 24 mrad case. The example CBED patterns in Figs. \ref{fig:bf_vs_df_combined}(a,b) make this concrete: most of the features remain visible at the lower dose for the 12 mrad case, whereas in the 24 mrad case the majority of the 4D STEM intensity information is dominated by noise.

Irrespective of resolution considerations, extending even a short way into the dark field region can be advantageous on experimental data. It is not uncommon to encounter detector defects such as hot or dead pixels, or random impacts from cosmic rays, that result in almost single-pixel localised deviations from the larger pattern. While we may counter these effects somewhat with additional preprocessing, they still represent sources of model mismatch in our reconstructions. Increasing the number of detector pixels included by extending the radius of the area of interest in diffraction space helps suppress any residual effect pixel defects might have.

There is of course a limit as to how much the dark-field signal might improve the quality of our reconstruction. As thickness increases, the positive contribution of thermally scattered electrons to the diffraction patterns becomes more pronounced and increases the signal in the dark-field region, partially manifesting in Kikuchi lines as seen in Figure \ref{fig:frozen_phonon_vs_abs}. Since the absorptive model discards those high-angle scattered electron and simply attenuates the signal in the diffraction patterns, this source of model mismatch increases with increasing thickness. While perhaps not the greatest limiting assumption in reconstructing samples beyond 500~{\AA} in thickness, it is nonetheless a contributing factor, one already observed by \citet{diederichs2024exact}.

\section{\LARGE Experimental reconstruction}\label{sec:exp_recon}

We now leverage all the improvements to the reconstruction algorithm descibed above to achieve structure determination from experimental data. Using an aberration-corrected Thermo Scientific Spectra $\upvarphi$, equipped with a CEOS SCORR+ probe Cs corrector and a high-brightness X-FEG source, we acquired 4D STEM data at 300 kV accelerating voltage, with a calibrated convergence semiangle of 21.2 mrad to ensure substantial disk overlap between Bragg beams. Care was taken to minimise aberrations, though, as we show presently, defocus remained appreciable and needed to be included in the reconstruction. The optical configuration of the microscope allowed us to fix the probe-forming condenser lenses (and thus the aberrations, including the focal condition) while demagnifying the source to improve coherence. We chose a demagnification at which source magnification ceased to be the dominant factor limiting spatial incoherence. The specimen was a wedge-polished, monolithic SrTiO$_3$ crystal oriented along the [001] zone axis. Although we endeavoured to align the sample on zone, a small degree of mistilt remained and so was allowed for in the reconstruction.\footnote{A minor modification to the work of \citet{sadri2023determining} allows us to include mistilt as a quantity to refine during the optimisation process.} The 4D STEM data was acquired with a Gatan K3 IS direct electron detector, using the proprietary correlated double sampling mode to reduce reset noise and diminish fixed pattern noise \citep{SunCDSGatan2021}, and suppressing inelastic scattering contributions by energy filtering via a Gatan Continuum 1069HR Imaging filter with an energy selection slit set to 10 eV in width and centred on the zero-loss peak. The probe was scanned over a 45~{\AA} $\times$ 45~{\AA} region with a probe step size of 0.1~{\AA} (approximately twice Nyquist sampling). A new gain reference was acquired prior to the scan and implemented manually afterwards, to increase the achievable frame rate to a dwell-time of 1.05 ms. With an average probe current of about 80 e$^-$/pixel/s, this corresponds to an incident dose of approximately 1.7 $\times$ 10$^6$ e$^-$/{\AA}$^2$. We used native cutting and binning on the K3 when collecting the diffraction data to reduce data size, with a reciprocal space sampling of 0.21 mrad/pixel across 512$\times$512 pixels.

Figure \ref{fig:exp_data_overview} overviews this experimental data. The virtual annular bright field (ABF) image in Fig. \ref{fig:exp_data_overview}(a) suggests the probe aberrations are reasonably balanced and that the defocus is at or near the sample surface. The latter can be deduced by the contrast displayed in the ABF image: both heavy and light atomic columns are visible and in good agreement with the expected lattice structure of SrTiO$_3$. Following previous work \citep{findlay2010dynamics}, such conditions suggest the beam waist of the focused electron probe is at or near the sample surface. The individual CBED patterns in Fig. \ref{fig:exp_data_overview}(b) contain structure clearly visible above noise (at least in the bright field region) and not washed out by spatial incoherence. The position averaged CBED (PACBED) pattern in Fig. \ref{fig:exp_data_overview}(c) indicates that the sample is reasonably thin and that some slight mistilt is present broadly in the negative k$_x$ and positive k$_y$ direction, as indicated by the arrow extending from the centre of the image.

\begin{figure*}[ht]
    \centering
    \includegraphics[width = \textwidth]{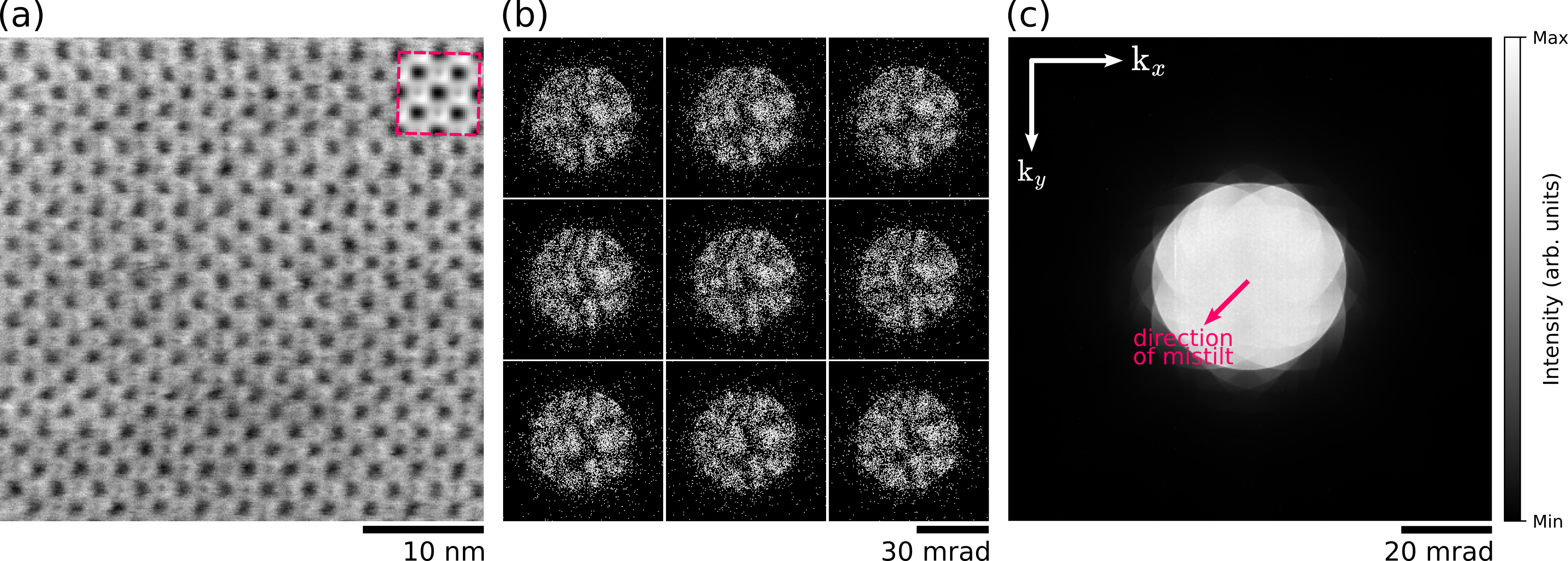}
    \caption{(a) Virtual ABF image from raw data, with the mean unit cell inset. (b) 3$\times$3 array of CBED patterns from the raw 4D STEM dataset, cropped for display purposes and with contrast enhanced to make variations across adjacent patterns more visible. (c) PACBED pattern of the pre-processed 4D STEM data prior to drift correction. The pink arrow indicates the direction (but significantly exaggerates the magnitude) of the mistilt.
    }
    \label{fig:exp_data_overview}
\end{figure*}

We performed standard pre-processing on the raw 4D STEM data, including applying a flat-field image to demodulate the detector response (by dividing each diffraction pattern in the array with a flat-field reference image that was acquired in a vacuum region), binning in reciprocal space by a factor of 4, and cropping reciprocal space beyond 32 mrad (thereby making some use of information from the dark field). Binning increases the signal-to-noise per pixel in the diffraction patterns, though, as per Fig. \ref{fig:frozen_phonon_vs_abs}(c), doing so can be a source of model mismatch impacting the accuracy of reconstructing high-order Fourier coefficients. Cropping lets us discard those dark-field scattered electrons for which the SNR is too low to provide useful constraints, with the added benefit of reducing the array size and computational load. We also corrected for the linear and non-linear scan distortions in the slow-scan direction evident in Fig. \ref{fig:exp_data_overview}(a) by using orthogonal scan pairs as outlined in \citet{ophus2016correcting}. This was followed by mean unit cell averaging across a region of common overlap between the two drift-corrected scans, converting random fluctuations in probe position into an increase in the effective spatial incoherence and reducing the computational cost by building in foreknowledge of the periodicity of the material. The mean unit cell of the virtual ABF image is inset in the top right corner of Fig. \ref{fig:exp_data_overview}(a). The mean unit cell diffraction data was then further binned by a factor of 3 in each linear direction in reciprocal space. The resulting data is a fraction of the size and contains fewer sources of model mismatch beyond those we typically expect such as partial spatial coherence, noise and unknown defocus.

Structure determination here means solving for the projected potential, but doing so reliably also requires us to determine the sample thickness, the defocus (assuming other aberrations are minimal), the effective source distribution (via some low order parameterisation), and sample mistilt. We do this via a limited grid search.\footnote{With judicious choice of metrics, the process might be made more efficient following the Bayesian optimisation procedure of \citet{cao2022automatic}, but grid search suffices here and the intermediate steps help bolster our understanding and interpretation.} Since the ABF image was reasonably in focus, we search in defocus from $90$~{\AA} overfocus to $-90$~{\AA} underfocus. In each case, we took the thickness from the $t+2\Delta f$ estimate predicted via antidiagonal symmetry. Since the PACBED indicated modest tilt, we vary the mistilt ${\bf k}_\perp$ in Eq. (\ref{eq:a_matrix}), re-expressed for ease of interpretation in scattering angle units of mrad, over the three combinations $(0,0)$ mrad, $(-2,2)$ mrad and $(-4,4)$ mrad. Trial and error reconstructions assuming a Gaussian effective source distribution failed (not shown), but those assuming a Lorentzian effective source distribution were more successful. The effective source distribution function $\Lambda({\bf R})$ was thus assumed to be purely Lorentzian, with HWHM values of 0.2~{\AA}, 0.4~{\AA}, and 0.6~{\AA} considered in the initial grid search. To make the grid search less time consuming, the reconstructions were halted after three iterations of reconstructing {\am}, with the apodisation function only extending to approximately half the convergence semiangle. By that point the low-order Fourier coefficients have largely ceased changing, and this suffices to tell whether or not the parameter combination appears promising.

\begin{figure*}[htp]
    \centering
    \includegraphics[width = 0.7\linewidth]{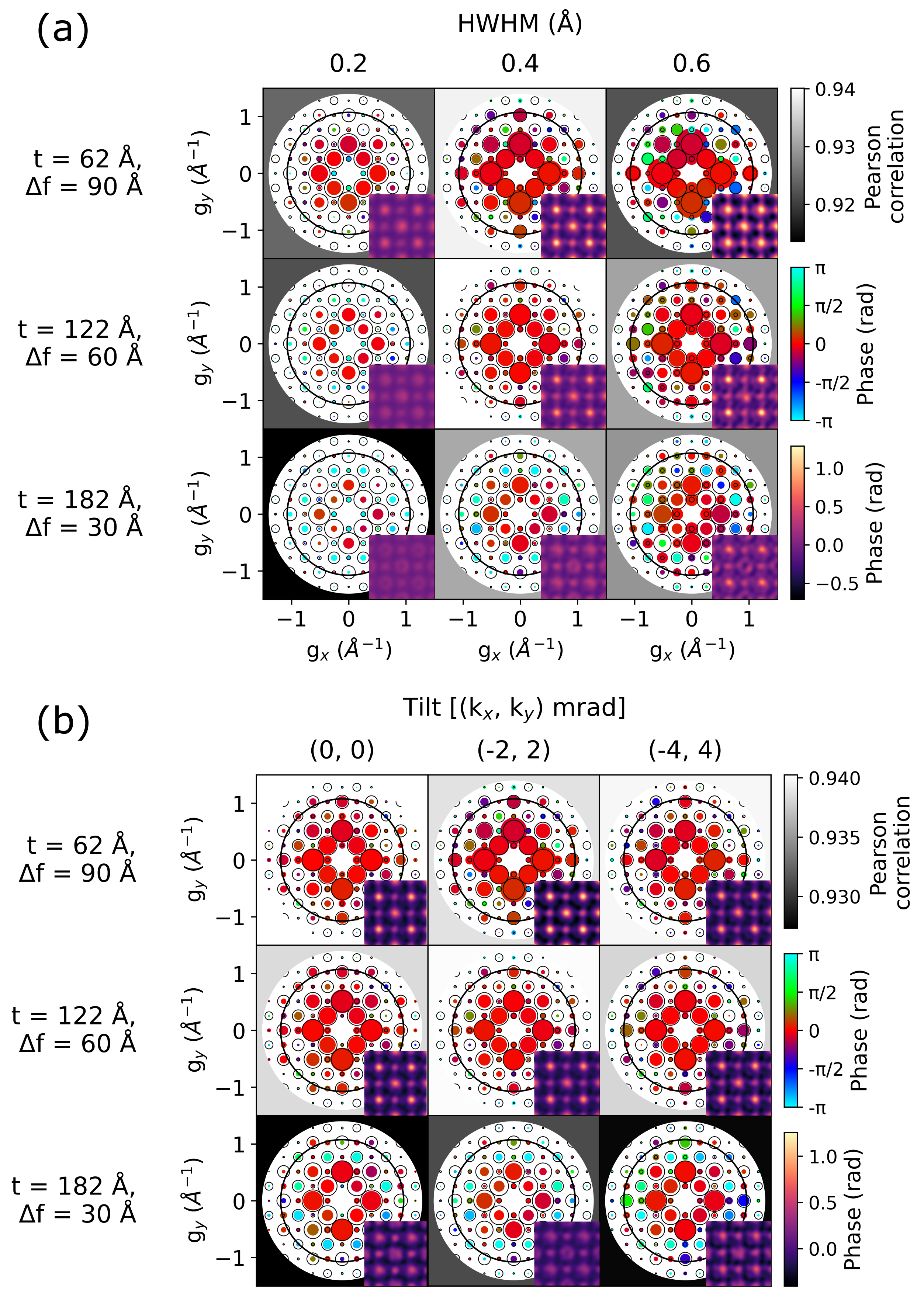}
    \caption{(a) Visualisation of the reconstructed potential from experimental 4D STEM SrTiO$_3$ data resulting from a grid-search over HWHM of the (assumed Lorentzian) effective source distribution (columns) and defocus values (rows) for a fixed mistilt of $(-2, 2)$ mrad. The starting guess for sample thickness is set via the antidiagonal symmetry constraint from the initial {\sm} reconstruction and the assumed defocus, and the reconstructions terminated early as described in the body text. (b) As in (a), but for a fixed 0.4~{\AA} HWHM effective source distribution while varying the sample mistilt (columns). The background colour around each plot shows the value of the Pearson correlation between the experimental and simulated 4D STEM data (greyscale colorbar), and tends to be higher for more accurate reconstructions.    
    }
    \label{fig:exp_am_search}
\end{figure*}

Figure \ref{fig:exp_am_search}(a) shows a subset of the grid search, specifically over effective source HWHM and a favourable range of defocus. Three measures of reconstruction quality are shown. The first measure is the Fourier coefficient visualisation. Though by far the most informative about reconstruction quality, this measure is only possible with foreknowledge of the SrTiO$_3$ projected potential. This would not be available for a sample of unknown structure, and so we should not estimate favourable parameter combinations by this measure. It is shown here only to establish that the other measures correlate with it. The second measure is the real-space projected potential visualisations, shown on a common scale. Assuming only a low order zone axis crystal, we regard as more favourable those parameter combinations for which these potential visualisations show clear atomistic structure and good contrast. For example, the parameter combination $\Delta f = 30$ {\AA}, ${\rm HWHM} = 0.6$ {\AA} is unfavourable because of the doughnut contrast of the central column and the parameter combination $\Delta f = 60$ {\AA}, ${\rm HWHM} = 0.2$ {\AA} is unfavourable because the contrast is very low. The third measure is the Pearson correlation between the experimental 4D STEM intensities and the 4D STEM intensities simulated from the retrieved structure, shown here via the background grayscale level surrounding each Fourier space plot. 

Fig. \ref{fig:exp_am_search}(b) shows a similar grid-search to that in Fig. \ref{fig:exp_am_search}(a), sweeping over the same range of defocus and thickness values, but assuming a fixed effective source distribution at 0.4~{\AA} HWHM and instead varying the sample mistilt. There appears to be some degree of robustness to mistilt, with the best result for each assumed mistilt (90~{\AA} defocus for $(0, 0)$ mrad and $(-4, 4)$ mrad, 60~{\AA} defocus for $(-2, 2)$ mrad) all having very similar Pearson correlations of about 0.94. In all three cases, the low-order Fourier coefficients in the coloured circle plots are well-converged, and the inset real-space potentials display reasonable contrast variation across the field of view, reflect the expected atomic structure of SrTiO$_3$, and do not exhibit phase reversal artefacts like doughnut contrast.

Figs. \ref{fig:exp_am_search}(a,b) show reasonable correlation between the three measures of reconstruction quality, with all suggesting ${\bf k}_\perp = (-2,2)$ mrad, $\Delta f = 60$ {\AA} and ${\rm HWHM} = 0.4$ {\AA} to be a promising parameter combination. The corresponding sample thickness estimate is 120~{\AA}, which is encouragingly consistent with the PACBED pattern in Fig. \ref{fig:exp_data_overview}(c). The estimated sample mistilt is also consistent with the approximately 3 mrad net mistilt that can be estimated from the PACBED pattern in Fig. \ref{fig:exp_data_overview}(c). Note that estimating thickness and tilt via PACBED is only possible when the structure is already known.

\begin{figure*}[htb]
    \centering
    \includegraphics[width = \textwidth]{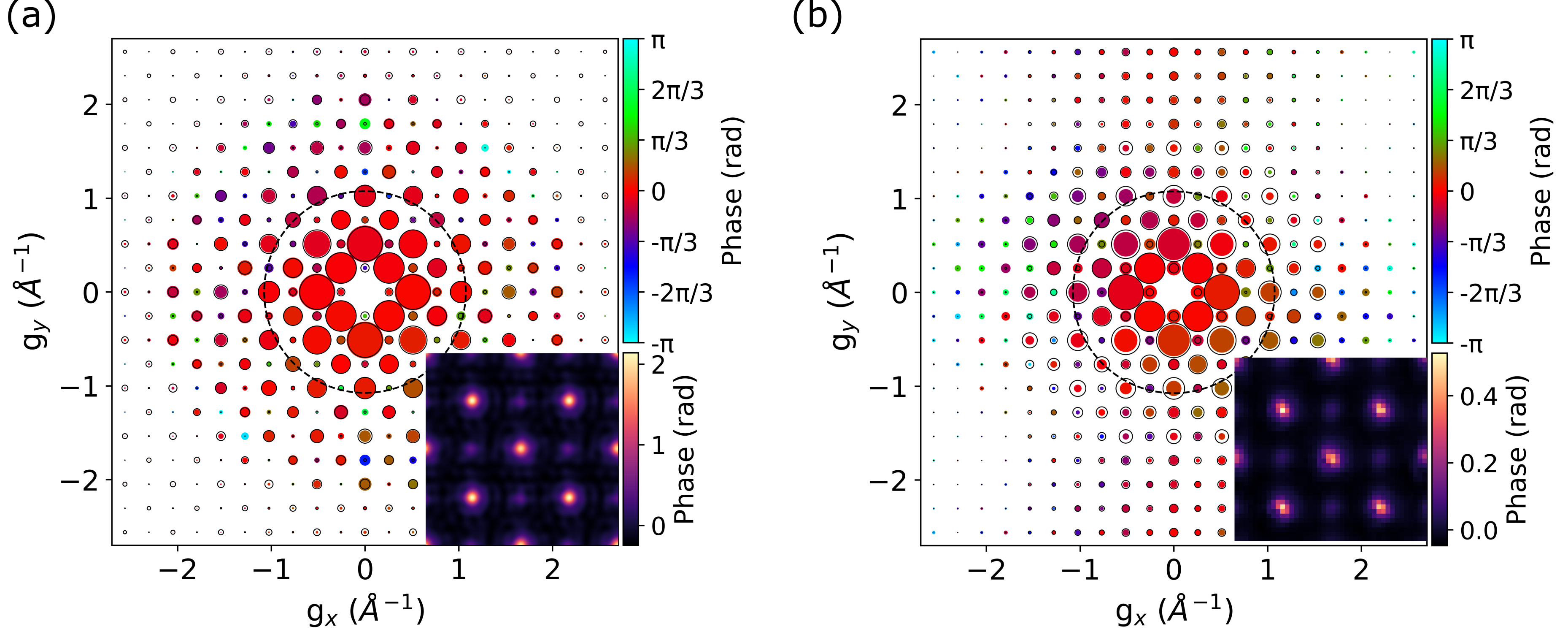}
    \caption{(a) Reconstructed elastic projected potential from experimental 4D STEM data via the {\sm} matrix. Defocus is assumed to be approximately 70~{\AA}, corresponding to a sample thickness of 110~{\AA}, with a Lorentzian spatial incoherence distribution function with an assumed HWHM of 0.3~{\AA}. (b) Single slice of the reconstructed elastic potential from mixed-state inverse multislice ptychography. Algorithm initialised using the same parameters as the {\sm} matrix/{\am} matrix reconstruction, with a probe defocus of 70~{\AA} and a sample thickness of 110~{\AA}. Spatial incoherence was accounted for by decomposing the electron probe into 6 mutually incoherent probe modes.}
    \label{fig:exp_am_recon}
\end{figure*}

We performed a subsequent, finer grid search centred on the promising parameter combinations identified in the initial, coarser grid search, this time using seven iterations of reconstructing {\am}, with the apodisation function extending to approximately twice the convergence semiangle. The final experimental reconstruction shown in Fig. \ref{fig:exp_am_recon}(a) assumes a Lorentzian effective source with HWHM 0.3~{\AA}, $\Delta f = 70$ {\AA} and thickness 110~{\AA}. The Fourier coefficients within the bright-field disk are accurately reconstructed in both phase and amplitude. The reconstruction itself is performed out to just beyond 2~{\AA}$^{-1}$, where our choice of apodisation function causes the experimental Fourier coefficients to terminate. Going out beyond the bright-field disk, we observed the expected gains in reconstruction quality when incorporating high-angle scattered electrons. The use of the dark field is effective in ensuring accurate Fourier coefficient amplitudes out to the full extent of {\am} allowed to vary. The expected SrTiO$_3$ structure, including the weakly-scattering oxygen columns that were not clearly visible in many earlier reconstructions from simulated data (due to the higher noise level in those simulations relative to that of the mean-unit-cell-averaged experimental data), are clearly visible in the reconstructed real-space projected potential: the atoms are localised at their expected coordinates, and only a trace level of noise artefacts is visible.

To round out our discussion of the {\sm} matrix reconstruction, it is useful to compare this approach with inverse multislice ptychography. As a concrete example, Fig. \ref{fig:exp_am_recon}(b) shows a reconstruction using mixed-state inverse multislice ptychography as implemented in \emph{py4DSTEM} \citep{savitzky2021py4dstem} (see \citet{varnavides2023iterative} for a detailed explanation of the methodology used in \emph{py4DSTEM}), which solves for arbitrary sample potentials on a pixel array. The similar quality of the {\sm} matrix reconstruction in Fig. \ref{fig:exp_am_recon}(a) and the inverse multislice ptychography reconstruction in Fig. \ref{fig:exp_am_recon}(b) is evident. We have made equivalent assumptions where possible, but it is perhaps the differences that are most informative.

The ptychographic reconstruction used the same pre-processed mean unit cell 4D STEM data (and removed the zero-padding used to handle unconstrained edges of the potential in \emph{py4DSTEM}). However, in order to adequately sample both real and reciprocal space in the multislice algorithm we did not apply the additional 3$\times$3 binning or cropping to 32 mrad. The inset potential in Fig. \ref{fig:exp_am_recon}(b) is shown at the resultant sampling, though the visible pixelation could be ameliorated by Fourier interpolation. By working entirely in Fourier space, the {\sm} matrix approach uses a naturally sparse basis well suited to periodic samples. This sparsity is not readily achieved for well-converged multislice calculations when solving for arbitrary sample potentials on a pixel array.\footnote{Recent work shows how considerable sparsity can be introduced to multislice ptychography by presupposing atomic potentials or basis states thereof \citep{yang2024local,diederichs2024exact}.} However, multislice ptychography can handle potentials non-periodic in the lateral direction with no increase in computational effort. The same cannot be said for the {\sm} matrix approach, where the computational complexity of the matrix-based calculation increases steeply when representing non-periodic object via a Fourier basis. Also, the (hopefully small) model mismatch likely introduced by the imperfect repeat unit averaging can be avoided in ptychography by using the non-averaged data and instead correcting for residual errors in probe position by probe position relaxation \citep{dwivedi2018lateral}.

For direct comparison, the ptychographic reconstruction has assumed the same approximately 110~{\AA} sample thickness determined in the {\sm} reconstruction\footnote{The approximately 110~{\AA} thick sample was decomposed into a series of 21 slices, each roughly 5.5~{\AA} thick, slightly thicker than an ideally converged multislice calculation, but nonetheless sufficiently thin that amplitude loss in each slice does not adversely impact the reconstruction.}. Similarly, the ptychographic reconstruction constrained the potentials in all slices to be identical. This constraint may readily be relaxed in ptychography (though some regularisation constraints may be necessary \citep{chen2021electron}) --- indeed, structures that vary along the beam direction are of great interest for materials science applications. There seems no logical impediment to a similar generalisation of the {\sm} matrix reconstruction approach using stacked scattering matrices \citep{pennington2014third}. However, given lack of periodicity in the lateral direction is a greater challenge for the {\sm} matrix approach and the class of specimens which might be periodic in the lateral direction but non-periodic along the optical axis is limited, it seems doubtful the {\sm} matrix approach would be competitive for substantially non-periodic objects. However, in samples where periodicity is expected, the {\sm} matrix approach offers advantages. The forward calculation time is independent of sample thickness in the {\sm} matrix formulation (whereas it scales linearly with thickness in the multislice formulation).\footnote{We are careful to say ``forward calculation'' here because the reconstruction time can depend on thickness not just through the scaling of the forward calculation time but also through the complexity of the 4D STEM data, which can affect the ``difficulty'' of the reconstruction process in a nonlinear way.} Also, as discussed earlier, if the probe is largely aberration free the symmetry properties of {\sm} may reduce the parameter space searching needed for thickness estimation.

Perhaps the biggest philosophical difference between the {\sm} matrix and ptychographic reconstructions shown in Fig. \ref{fig:exp_am_recon} is in modelling the incident wavefield. In the latter, we used mixed-state or multi-mode ptychography \citep{thibault2013reconstructing}, whereby the incident wavefield was modelled as a set of mutually incoherent wavefields. The 6 probe modes determined for the reconstruction in Fig. \ref{fig:exp_am_recon}(b) are shown in Fig. \ref{fig:ptychography_reconstructed_probe_modes} in the Appendix. The first mode seems quite plausible, with its phase profile showing some defocus plus residual astigmatism. The remaining modes are difficult to interpret. Given the decomposition of a density matrix into incoherent modes is not unique, the density matrix associated with these modes might be plausible even if the particular decomposition returned by the reconstruction algorithm is not intuitive. However, it is also possible that these modes are simply absorbing model mismatch without being innately meaningful. Such an explanation seems likely when one considers the acquisition parameters for this experiment were not optimised for ptychography (such as those laid out in \citet{gilgenbach2024methodology}) and the difficulty encountered to achieve the reconstruction shown in Fig. \ref{fig:exp_am_recon}(b). Additional constraints or regularisation can be imposed on the probe modes to ensure they are more physical, however explorations (not shown) indicated that doing so came at the expense of reconstruction accuracy, with errors that may have previously been absorbed by said probe modes contaminating the object and thus degrading the quality of the retrieved electrostatic potential. An alternative, as used here in the {\sm} matrix reconstruction (though the strategy could be used in optimisation-based ptychographic reconstructions), is to model coherent aberrations and spatial incoherence with physics-based models with relatively few degrees of freedom, here defocus and the width of the assumed Lorentzian effective source distribution. We find the greater interpretability of this approach attractive, though if the parameterisation used is not general enough to encompass the coherent and incoherent aberrations present then that would produce some model mismatch.

Both the {\sm} matrix and inverse multislice ptychography reconstructions account for thermal diffuse scattering phenomenologically via an absorptive potential. \citet{diederichs2024exact} show that the intensity contribution from thermally scattered electrons can be reproduced in inverse multislice ptychography if one assumes the atomic potentials and includes the thermal vibrational amplitude as an optimisation parameter with a frozen phonon approach to the forwards multislice calculation. \citet{gladyshev2023reconstructing} suggest a mixed state approach for the sample might sidestep the need to assume the atomic potentials. That said, in the present case assuming an absorptive potential does not appear to appreciably limit the reconstruction, though the reconstructed absorptive potential is not in good quantitative agreement with simulation. 

In summary, both through the nature of the multislice method and through the increasing availability of implementations, inverse multislice ptychography seems better suited to determining 3D structure in non-periodic samples. However, for periodic samples, the {\sm} matrix approach has much to recommend it, including computational complexity independent of thickness, symmetry properties that allow for consistency checks and estimation of parameters like thickness and defocus, and the possibility of interrogating the sample with a reconstructed {\sm} matrix without necessarily taking the additional step of solving fully for the structure.

\section{\LARGE Conclusion}
We have presented three improvements to the scattering matrix method for determining the projected electrostatic potential from 4D STEM data of thick samples where multiple electron scattering is significant.

First, we showed how to modify the phase retrieval step to account for spatial incoherence. In simulation, the improvement this afforded was such that little-to-no further advantage was gained by using 4D STEM data from two different defocus values. Overestimating the width of the effective source distribution allowed reliable reconstructions whereas underestimating it did not.

Second, we showed that in the presence of unknown defocus, enforcing the anti-diagonal symmetry property of {\sm} allows an estimate of the sample thickness coupled with the probe defocus (specifically, $t+2\Delta f$). High dose data produced estimates accurate to within a couple of nanometers, though the variability of the estimate increases when the dose decreases and the initially-reconstructed {\sm} becomes less accurate.

Third, we showed how to extend the procedure to use electrons scattered beyond the bright-field disk. Though hampered somewhat by binning artefacts, such dark-field signal can be used to enhance the fidelity of the reconstructions, and potentially allow for the use of smaller probe forming apertures than previously realised. Ultimately, the low signal-to-noise ratio per pixel at higher scattering angles places limits on the spatial frequencies that can reliably be reconstructed.

All these advances were leveraged to reconstruct the projected potential of a sample of monolithic SrTiO$_3$ from experimental 4D STEM data. This required reasonable starting estimates for thickness, defocus, effective source size and mistilt, which were found by a grid search strategy drawing on the improvements above and using both qualitative (appearance of the reconstructed real-space potential) and quantitative (correlation between the experimental and simulated 4D STEM data) measures to identify favourable values for these quantities. The result was a reconstruction in excellent quantitative agreement with the expected structure.

\begin{appendices}

\section{\Large Appendix}\label{sec:appendix}
The results presented in Fig. \ref{fig:exp_am_recon}(b) show the elastic potential reconstructed by mixed-state inverse multislice ptychography. One of the strengths of ptychography lies in its ability to simultaneously solve for the probe. In the case of mixed-state ptychography, mutually incoherent probe modes can be used to account for partial coherence. Figure \ref{fig:ptychography_reconstructed_probe_modes} shows the six reconstructed probe modes that accompany the object reconstructed in Fig. \ref{fig:exp_am_recon}(b) and their relative contributions to the overall probe intensity.

As discussed in the main body text, the first or primary probe mode is the most interpretable and contributes the most to the overall intensity at 32.8\%. Its profile in both real and reciprocal space is suggestive of residual aberrations beyond defocus, most likely astigmatism. The other probe modes are less informative on inspection, though likely act as a sink for errors in the reconstruction.

\begin{figure}[htb!]
    \centering
    \includegraphics[width=\linewidth]{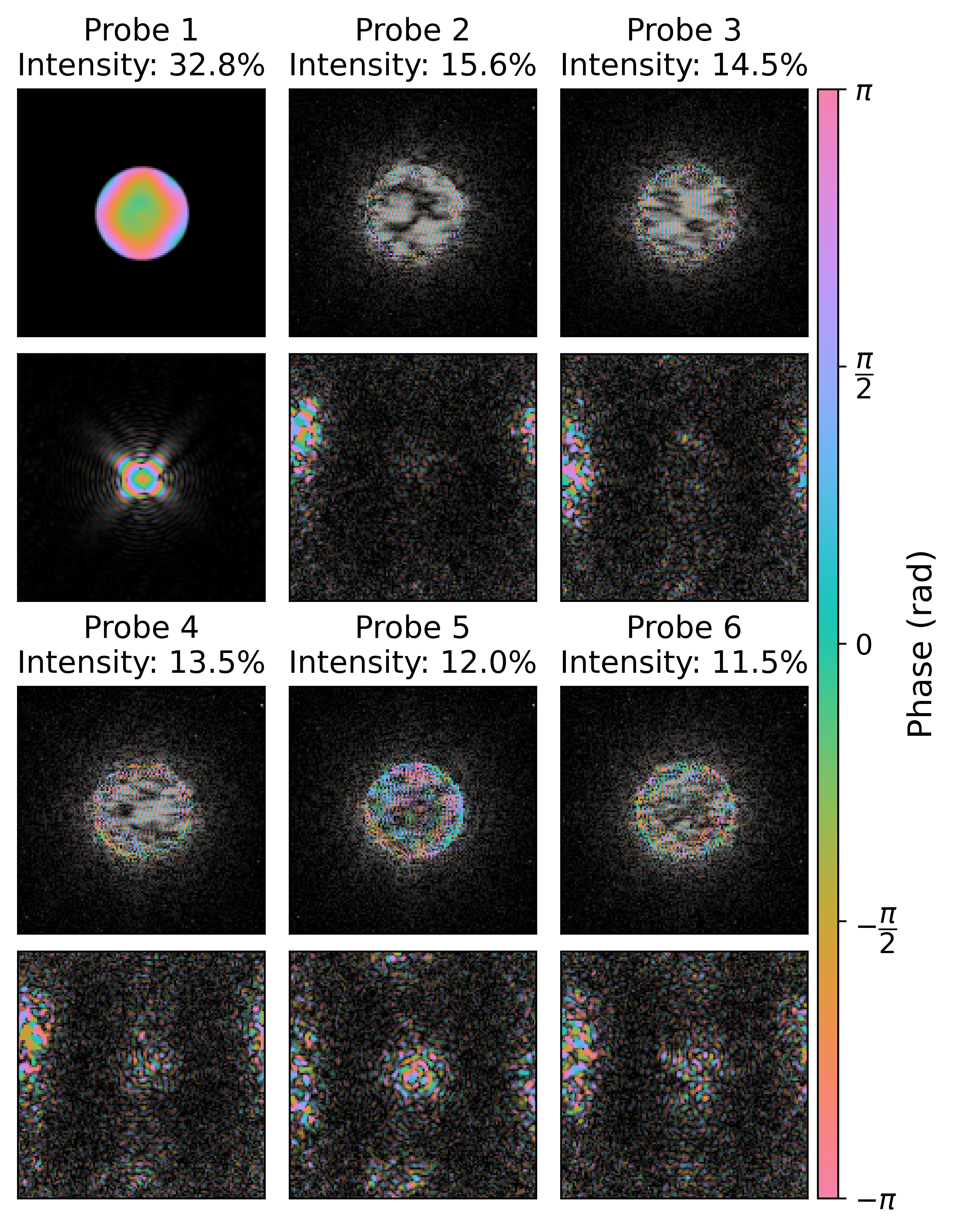}
    \caption{Reconstructed probe modes from mixed-state inverse multislice ptychography. The incident electron wavefunction has been decomposed into six mutually incoherent states to handle partial coherence. Each probe mode is shown in both reciprocal-space (top) and real-space (bottom), with its relative contribution to the overall intensity also displayed.}
    \label{fig:ptychography_reconstructed_probe_modes}
\end{figure}

\end{appendices}

\section{\Large Competing interests}
No competing interest is declared.

\section{\Large Author contributions statement}
E.W.C.T.-L., C.O., T.C.P., and S.D.F. conceived the study.
E.W.C.T.-L., M.W., and T.C.P. conducted the experiments.
L.B. and M.W provided specimens.
E.W.C.T.-L., A.S. and C.O. analysed the results.
S.M.R., G.V., C.O. and E.W.C.T.-L. performed the ptychography analysis. 
E.W.C.T.-L., T.C.P. and S.D.F. wrote the manuscript.
All authors contributed to discussion and editing the manuscript.

\section{\Large Acknowledgments}
This research is supported by an Australian Government Research Training Program (RTP) Scholarship. SDF is the recipient of an Australian Research Council Future Fellowship (project number FT190100619) funded by the Australian Government. Work at the Molecular Foundry was supported by the Office of Science, Office of Basic Energy Sciences, of the U.S. Department of Energy under Contract No. DE-AC02-05CH11231. The authors acknowledge the use of the instruments and scientific and technical assistance at the Monash Centre for Electron Microscopy, a Node of Microscopy Australia. This research used equipment funded by Australian Research Council grant LE170100118.

\bibliographystyle{abbrvnat}
\bibliography{parallax1}

\end{large}

\end{document}